\documentclass[12pt]{article}
\usepackage{graphicx}
\usepackage{subcaption}
\usepackage{amsfonts}
\usepackage{amssymb}
\usepackage{float}
\usepackage{caption} 
\usepackage{mathrsfs}
\usepackage{amsmath}
\usepackage{indentfirst}
\usepackage{hyperref}
\usepackage{cite}
\usepackage{microtype}

\usepackage{authblk}

\begin{document}
\renewcommand{\thefootnote}{\fnsymbol{footnote}}
\title{Shadow and Polarized Images of Schwarzschild-MOG Black Hole Illuminated by Thick Accretion Disk}
\author[1]{\small Cheng Hu}
\author[1]{\small Huan Ye}
\author[2]{\small Hong Lei}
\author[3]{\small Zi-Chao Lin\thanks{corremail address: linzch24@cqu.edu.cn}}

\affil[1]{\small Chengdu Neusoft University, Chengdu 610000, China}
\affil[2]{\small Chongqing Bashu Science City Secondary School, Chongqing 401331, China}
\affil[3]{\small College of Physics, Chongqing University, Chongqing 401331, China}

\date{}
\maketitle
\begin{abstract}
Recent EHT polarimetric observations provide a powerful probe of near-horizon physics and modified gravity. We investigate the shadow and polarization images of a static Schwarzschild-MOG black hole illuminated by geometrically thick, optically thin accretion flows. Adopting a phenomenological model and the Hou disk model, and solving null geodesics and relativistic radiative transfer, we study how the MOG parameter, inclination angle, anisotropic synchrotron emission, and flow geometry affect the total intensity and linear polarization. It is found that the MOG parameter mainly enlarges and brightens higher-order photon rings and broadens the shadow, while the inclination angle breaks axial symmetry and enhances polarized emission inside the shadow. The Hou disk model suppresses off-equatorial radiative interference and localizes polarization signals within higher-order rings. Electric vector position angle (EVPA) patterns are mostly azimuthal but vary rapidly near lensed images. These results provide theoretical diagnostics for constraining MOG with EHT/ngEHT polarimetric observations.
\setlength{\parindent}{0pt} \setlength{\parskip}{1.5ex plus 0.5ex minus 0.2ex} 


\end{abstract}

\thispagestyle{empty}

\newpage

\section{Introduction}

Within the framework of General Relativity (GR), black holes are spacetime singularities resulting from extreme gravitational collapse, shielded by an event horizon that acts as a causal one-way membrane~\cite{oppenheimer1939continued, penrose1965gravitational, hawking1973large}. This horizon precludes the escape of matter and radiation, governing mass accretion in the classical paradigm. In recent years, the multi-messenger era has transitioned black hole physics from theoretical postulation to precision observational science. Milestones such as gravitational wave detections by LIGO/Virgo~\cite{abbott2016observation} and horizon-scale imaging of M$87^*$ and Sgr A$^*$ by the Event Horizon Telescope (EHT)~\cite{akiyama2019firstm87, akiyama2022firstsgr} provide unprecedented evidence for black holes. Furthermore, follow-up EHT polarimetric observations have elucidated magnetized plasma evolution within accretion flows~\cite{akiyama2021firstm87vii, akiyama2021firstm87viii, akiyama2024firstsgrvii, akiyama2024firstsgrviii}. These polarized images offer a novel window for probing near-horizon radiation mechanisms and magnetic topologies, while providing a stringent testing ground for GR and alternative theories of gravity in the strong-field regime~\cite{psaltis2020gravitational, kocherlakota2021constraints}.

Despite its success in the weak-field regime, GR faces major challenges on both astrophysical and cosmological scales. On cosmological scales, standard GR cannot naturally account for late-time cosmic acceleration \cite{Riess1998, Perlmutter1999, astier2012observational}, flat galactic rotation curves \cite{rubin1970rotation, rubin1980rotational, moffat2013mog}, or persistent tensions such as the $H_0$ and $S_8$ discrepancies, without introducing ad hoc dark sectors \cite{ade2016planck, Planck2020, DiValentino2021}. On astrophysical scales, the mass discrepancy in galaxy clusters, where dynamical mass greatly exceeds baryonic luminous mass \cite{zwicky1937masses, trimble1987existence}, is typically addressed by invoking dark matter and dark energy, yet neither has achieved direct, non-gravitational detection \cite{riess2004type, Copeland2006, Bertone2005, astier2006supernova, Arcadi2018}. Theoretically, the inevitability of spacetime singularities and the lack of a complete quantum gravity framework further expose the limits of GR. These unresolved issues have motivated the exploration of modified theories of gravity as a vital frontier in modern gravitational physics.

To address these discrepancies, modern gravity research generally follows two paths: introducing non-baryonic dark matter, or modifying the long-range behavior of gravity via generalized geometry and additional degrees of freedom \cite{Clifton2012, Nojiri2017}. Consequently, numerous modified gravity models have been proposed, including higher-curvature Gauss-Bonnet gravity \cite{Glavan2020}, $f(R)$ gravity \cite{Sotiriou2010}, $f(R,T)$ gravity \cite{Harko2011}, extended scalar-tensor theories \cite{Fujii2003}, and teleparallel gravity \cite{Cai2016}. Among these alternatives, Scalar-Tensor-Vector Gravity (STVG), also widely known as MOdified Gravity (MOG), formulated by John Moffat \cite{Moffat2006}, provides a compelling covariant framework. Based on the principle of least action, MOG extends the Einstein-Hilbert action by introducing a massive neutral vector field and a scalar field that determines the dynamical gravitational coupling. This setup effectively enhances gravity on galactic and cosmological scales, successfully explaining flat rotation curves \cite{Brownstein2006}, cluster dynamics and gravitational lensing \cite{Moffat2014}, and the cosmic microwave background acoustic peaks \cite{Moffat2013CMB} without requiring non-baryonic dark matter \cite{Moffat2009, Moffat2013Galaxies}.

In the strong-field regime, the Schwarzschild-MOG spacetime serves as the standard static, spherically symmetric vacuum solution in STVG, corresponding to the non-rotating limit of the Kerr-MOG geometry \cite{Moffat:2014aja, Lima2020KerrMOG}.  Governed by the MOG parameter, the metric departs from the standard Schwarzschild solution: it features an effective gravitational charge that alters the event horizon radius, shrinks the photon sphere \cite{Sheykhi2018}, and shapes timelike and null geodesics \cite{Sharif2017}. Numerous studies have investigated the properties of Schwarzschild-MOG black holes, such as quasinormal modes and ringdown signals \cite{Manfredi2018, Liang2022}, thermodynamics and phase transitions \cite{Mureika2016}, particle acceleration via the B\'anados–Silk–West mechanism \cite{Pradhan2017}, and optical shadows under idealized toy-model illuminations \cite{Moffat2015Shadow, Guo2021}. However, synthetic polarimetric imaging of Schwarzschild-MOG black holes illuminated by realistic, geometrically thick and optically thin accretion flows remains largely unexplored \cite{Abramowicz2013, Narayan1995}. In particular, the interplay among the MOG parameter, relativistic frame effects, synchrotron emission, and magnetic field topologies (and how these factors shape distinct polarization signatures) has yet to be quantified~\cite{akiyama2021firstm87vii, akiyama2024firstsgrvii}. So it is essential to use high-precision horizon-scale polarimetry from the EHT and its next-generation successor (ngEHT) \cite{EHT2022SgrATest, Johnson2023ngEHT} for placing observational constraints on the MOG parameter.

Polarimetric imaging is a powerful tool for probing plasma dynamics and magnetic field configurations near black holes \cite{broderick2009imaging}. Polarized images of M$87^*$ and Sgr A$^*$, published by the EHT collaboration, reveal prominent polarization signatures across the emission ring, where the vectors encode the intrinsic magnetic geometry and order within the emission region \cite{akiyama2021firstm87vii, akiyama2024firstsgrvii}. These measurements exhibit distinct spiral-like patterns of electric vectors along the photon ring. Such electric vector position angle (EVPA) swings arise from the parallel transport of polarization vectors as polarized photons propagate through the strongly curved spacetime \cite{connors1980polarization}. By comparing synthetic polarized images with observations, one can constrain the astrophysical properties and geometry of black hole accretion flows \cite{akiyama2021firstm87vii, akiyama2021}. Motivated by these results, the EHT collaboration has successfully reconstructed the observed EVPA distribution and relative polarization intensity for M$87^*$ \cite{Narayan2021}. Building upon this framework, Gelles {\em et al}. applied Beloborodov's ray-tracing approximation in Schwarzschild spacetime \cite{Beloborodov2002, Gelles2021a} to model equatorial emitters and compute polarized images for Kerr black holes \cite{Gelles2021b}. Their work shows that polarization observables are jointly governed by the magnetic field geometry, black hole parameters, and viewing inclination. Furthermore, extensive studies have explored polarized imaging in alternative black hole spacetimes and horizonless ultracompact objects, offering promising ways to probe accretion physics and constrain underlying spacetime geometries \cite{deliyski2023polarized, Qin2022, yang2026shadow, zeng2026impact, zeng2025observational, chen2025polarized, shi2024polarized}.

Selecting an appropriate accretion disk model is crucial for the accurate synthetic black hole imaging. While classical thin-disk models offer convenient analytical benchmarks \cite{NovikovThorne1973}, they fail to capture the complex spatial distributions of plasma density and magnetic fields in hot near-horizon plasmas, prompting the widespread use of geometrically thick disk models \cite{Narayan1995}. Two primary theoretical frameworks are widely adopted to describe geometrically thick, optically thin accretion flows. The first is the standard phenomenological model, which relies on vertically averaged power-law profiles for electron number density and temperature \cite{moscibrodzka2016general, pu2016odyssey, yuan2003nonthermal, broderick2011evidence}. This model provides concise and reliable descriptions for large-scale accretion behavior at moderate accretion rates, enabling extensive parameter space studies. The second is the Hou disk model, a recently developed analytical solution derived from general relativistic magnetohydrodynamics (GRMHD) simulations \cite{Hou2024, Zhang2024}. By assuming that gravity dominates fluid acceleration near the event horizon, this model self-consistently captures the physical properties of both the thick disk and the jet funnel, rendering it an ideal framework for generating high-fidelity polarized images near the horizon.

Motivated by these observational and theoretical advances, this work systematically investigates the polarimetric signatures of a static Schwarzschild-MOG black hole illuminated by two distinct geometrically thick accretion disk configurations: the phenomenological model and the Hou disk model. Using these accretion flow models, we analyze how the polarization intensity and EVPA distribution vary with flow configurations, the MOG parameter, and the observer's inclination angle. Special attention is paid to polarization distortions induced by anisotropic synchrotron emission within geometrically thick disks. By comparing the polarimetric observables predicted by the two disk models, this study provides reliable theoretical diagnostics for testing modified gravity and constraining deviations from general relativity through future high-resolution polarimetric observations.

The paper is organized as follows. In Sec.~\ref{sec:2}, we briefly review the fundamental properties of the Schwarzschild-MOG black hole spacetime and derive the corresponding geodesic equations as well as the photon sphere. In Sec.~\ref{sec:3}, we formulate the theoretical framework of relativistic radiative transfer and thermal synchrotron emission, followed by a discussion on two geometrically thick accretion models. In Sec.~\ref{sec:4}, we investigate the unpolarized shadows and images of the black hole under both phenomenological and Hou disk accretion scenarios, analyzing their respective observational signatures. In Sec.~\ref{sec:5}, we illustrate the theoretical background for polarized radiative transfer under anisotropic emission in the Hou disk model and present the corresponding numerical results. Finally, we summarize and discuss our results in Sec.~\ref{sec:6}.

\section{Schwarzschild-MOG black hole and null geodesics}\label{sec:2}

\subsection{Review of black holes in MOG}
MOdified Gravity (MOG) extends the framework of general relativity by introducing three additional dynamical fields: a massive Proca-type vector field $\phi_{\mu}$, a scalar field $G(x)$ that promotes the Newtonian gravitational constant to a dynamical gravitational coupling, and a scalar field $\mu(x)$ associated with the vector field's mass\cite{Moffat2006}. The total action of this theory is constructed as the sum of its respective components, denoted by
\begin{equation}
S=S_G+S_\phi+S_S+S_M,
\end{equation}
where the gravitational action is given by
\begin{equation}
S_{G}=\frac{1}{16\pi}\int\frac{1}{G}(R-2\Lambda)\sqrt{-g}\,d^{4}x.
\end{equation}
For convenience, we set the speed of light $c$ and the standard Newtonian gravitational constant $G_N$ to unity. Here, the scalar field $G=G_N(1+\alpha)$ plays the role of a dynamical Newtonian gravitational constant, and $\Lambda$ is the cosmological constant. The dimensionless parameter $\alpha$  is commonly referred to as the MOG parameter. The vector field action is defined as
\begin{equation}
S_{\phi}=-\frac{1}{4\pi}\int\left[\mathcal{K}+V(\phi)\right]\sqrt{-g}\,d^{4}x,
\end{equation}
where the kinetic term $4\mathcal{K}=B^{\mu\nu}B_{\mu\nu}$ employs the Faraday-like tensor $B_{\mu\nu}=\partial_{\mu}\phi_{\nu}-\partial_{\nu}\phi_{\mu}$. Furthermore, the dynamics of the scalar fields are governed by
\begin{equation}
S_{S}=\int\frac{1}{G}\bigg[-\frac{1}{2}g^{\alpha\beta}\left(\frac{\nabla_{\alpha}G\nabla_{\beta}G}{G^{2}}+\frac{\nabla_{\alpha}\mu\nabla_{\beta}\mu}{\mu^2}\right)-\frac{V_{G}(G)}{G^{2}}-\frac{V_{\mu}(\mu)}{\mu^{2}}\bigg]\sqrt{-g}\,d^{4}x.
\end{equation}
Within this framework, the covariant current density is defined as
\begin{equation}
J^\mu=\kappa T_M^{\mu\nu}u_{\nu},
\end{equation}
where the energy-momentum tensor of a perfect fluid is expressed as
\begin{equation}
T_{M}^{\mu\nu}=(\rho_M-p_M)u^\mu u^\nu-p_M g^{\mu\nu}.
\end{equation}
Here, $u^\mu$ is the timelike four-velocity, and the coupling parameter $\kappa=\sqrt{\alpha G_{N}}$ incorporates the deviation parameter $\alpha=(G-G_{N})/G_{N}$ that quantifies the departure from Newtonian gravity. For a pressureless matter distribution ($p_M=0$), the current density elegantly simplifies to
\begin{equation}
J^\mu=-\kappa\rho_M u^\mu.
\end{equation}
In an asymptotically flat spacetime ($\Lambda=0$) devoid of matter ($T_{M}^{\mu\nu}=0$) and scalar fields, the generalized field equations reduce to
\begin{equation}
G_{\mu\nu}=8\pi G T_{\mu\nu}^{\phi},
\end{equation}
where $T_{\mu\nu}^{\phi}$ denotes the energy-momentum tensor of the massive vector field. In the static and spherically symmetric spacetime, the  metric has the standard form
\begin{equation}
ds^2=-f(r)dt^2+\frac{1}{f(r)}dr^2+r^2(d\theta^2+\sin^2\theta\,d\phi^2).
\end{equation}
By solving the corresponding vacuum field equations, the metric function is derived analytically as~\cite{Moffat:2014aja,Boboqambarova:2021cbf}
\begin{equation}
f(r)=1-\frac{2(1+\alpha)M}{r}+\frac{\alpha(1+\alpha)M^2}{r^2},
\end{equation}
where  $M$ represents the mass of the black hole. With this analytical solution, the fundamental theoretical basis is established for investigating black hole configurations and subsequent astrophysical phenomena within the MOG framework.

\subsection{Geodesic equations and effective potential}
To investigate the shadow and accretion disk profiles of a black hole in the Schwarzschild-MOG framework, we examine the motion of test particles (for both massive and massless) using the Lagrangian formalism. The line element for the Schwarzschild-MOG spacetime is governed by the metric function $f(r)$, and the corresponding Lagrangian $\mathcal{L}$ for a test particle of mass $m$ is given by
\begin{equation}
\mathcal{L}=\frac{1}{2}g_{\mu\nu}\dot{x^\mu}\dot{x^\nu}=\frac{1}{2}
\Big[-f(r)\dot{t}^2 + \frac{1}{f(r)}\dot{r}^2 + r^2\big(\dot{\theta}^2 + \sin^2\theta\,\dot{\phi}^2\big)\Big],
\end{equation}
where the overdot denotes differentiation with respect to the affine parameter $\lambda$. Due to the symmetries associated with the spacetime Killing vector fields, we identify two conserved quantities of motion: the specific energy $E$ and the specific angular momentum $L$, defined respectively as:
\begin{equation}
E \equiv -p_t = f(r)\dot{t}, \quad\quad L \equiv p_\phi = r^2 \sin^2\theta \, \dot{\phi}.
\end{equation}
Restricting our analysis to the equatorial plane without loss of generality ($\theta = \pi/2, \dot{\theta} = 0$), the normalization condition for the four-velocity $g_{\mu\nu}\dot{x}^\mu \dot{x}^\nu = -\kappa$ (where $\kappa = 1$ for massive particles and $\kappa = 0$ for photons) yields the fundamental radial equation:
\begin{equation}
-f(r)\dot{t}^2 + f(r)^{-1}\dot{r}^2 + r^2\dot{\phi}^2 = -\kappa.
\end{equation}
Substituting the expressions for the conserved momenta into the normalization condition leads to 
\begin{equation}
-f(r)\left(\frac{E}{f(r)}\right)^2 + f(r)^{-1}\dot{r}^2 + r^2\left(\frac{L}{r^2}\right)^2 = -\kappa.
\end{equation}
Rearranging the expression for $\dot{r}^2$, we obtain the general radial geodesic equation:
\begin{equation}
\dot{r}^2 = E^2 - V_{\text{eff}}(r),
\end{equation}
where the effective potential $V_{\text{eff}}(r)$ governing the radial motion is explicitly defined as
\begin{equation}
V_{\text{eff}}(r) = \Big[ 1 - \frac{2(1+\alpha)M}{r} + \frac{ \alpha(1+\alpha)M^2}{r^2} \Big] \Big( \kappa + \frac{L^2}{r^2} \Big).
\end{equation}
For massless photons ($\kappa = 0$), the radial equation simplifies to
\begin{equation}
\dot{r}^2 = E^2 - \frac{f(r)}{r^2}L^2.
\end{equation}
The impact parameter is defined as $b \equiv L/E$, representing the perpendicular distance of the photon trajectory from the optical axis. The boundary of the black hole shadow is determined by photons trapped in unstable circular orbits, which form the photon sphere. For a photon to maintain a stable or marginally stable circular orbit at a constant radius $r_p$, it must reside precisely at the peak of the effective potential barrier. This configuration then requires a vanishing radial velocity and the marginal gravitational instability. In other words, there should be a  turning point of the motion,
\begin{equation}
   V_{\text{eff}}(r_p) = E^2 \quad \Longrightarrow \quad \dot{r}^2 = 0,
\end{equation}
and the first derivative of the effective potential must vanish, with the second derivative being negative to ensure a local maximum (instability),
\begin{equation}
   V'_{\text{eff}}(r_p) = 0, \quad\quad V''_{\text{eff}}(r_p) < 0.
\end{equation}
These conditions uniquely determine the radius of the photon sphere $r_p$ and subsequently define the critical shadow radius observable by a distant observer.

\section{Preliminary and method}\label{sec:3}
In this section, we focus on the magnetized plasma environment near black holes and systematically elaborate the physical mechanism of electron synchrotron radiation, together with two distinct thick accretion disk models. We derive the covariant radiative transfer equation and its solution in the CGS unit system, and present both analytical and fitting forms for the synchrotron emissivity and absorption coefficient of ultrarelativistic electrons. In addition, we introduce the key structural parameters of the phenomenological accretion disk model and the Hou disk model, including electron number density, temperature, magnetic field, and fluid four-velocity. This section illustrates the theoretical and modeling foundation for subsequent radiative transfer calculations related to black hole imaging.

\subsection{Electron radiative}\label{Electron Radiative}
In this work, we account for the magnetic field in the vicinity of the black hole, where ultrarelativistic electrons within the plasma emit synchrotron radiation driven by the Lorentz force. For thick accretion disk models, key parameters including the particle number density, electron temperature, and magnetic‑field configuration play critical roles. In principle, these quantities can be derived by solving the general‑relativistic magnetohydrodynamic (GRMHD) equations. Nevertheless, owing to the substantial analytical complexity of the full GRMHD system, in this study we employ numerical methods together with reasonable simplifications.

\subsubsection{Radiative transfer equation}
The covariant form of the radiative transfer equation for unpolarized radiation is given by~\cite{gold2020verification}
\begin{equation}
\frac{d}{d\omega}\mathcal{I} = \mathcal{J} - \alpha \mathcal{I},\label{eq_dI}
\end{equation}
where $\mathcal{I}$, $\mathcal{J}$, and $\alpha$ are generalized invariants, and their relations to the actual physical quantities obey
\begin{equation}
 \mathcal{I} = \frac{I_\nu}{\nu^3}, \quad  \mathcal{J} = \frac{j_\nu}{\nu^2}, \quad \alpha = \nu\alpha_\nu.\label{eq_I_J_alpha}
\end{equation}
Here $I_\nu$ is the specific intensity defined as
\begin{equation}
I_{\nu} = \frac{dE}{dA\,dt\,d\Omega\,d\nu} = h^{4}\nu^{3}\frac{dN}{d^{3}r\,d^{3}p}.
\end{equation}
Its physical meaning is the energy $dE$ transported by electromagnetic waves within time $dt$, passing through an area $dA$, into a solid angle $d\Omega$, and over a frequency interval $d\nu$. Here we have used $E = N h \nu$ with $N$ denoting the number of photons. We further define
\begin{equation}
f = \frac{dN}{d^3 r d^3 p}
\end{equation}
as the number density distribution function of photons in phase space $(r, p)$. Since both $N$ and $d^3 rd^3 p$ are generalized invariants, the distribution function is then generalized invariant. In Eq.~\eqref{eq_I_J_alpha}, the emissivity $j_\nu$ is defined as
\begin{equation}
j_\nu = \frac{dE}{d^3 r dt d\nu d\Omega} = h^4 \nu^3 \frac{dN}{dt d^3 r d^3 p},
\end{equation}
and the absorption coefficient $\alpha_\nu$ obeys
\begin{equation}
\alpha_\nu = n \sigma_\nu,
\end{equation}
where $\sigma_\nu$ is the photon absorption cross-section, and $n$ is the number density of the medium. To convert from geometric units to CGS units, one may introduce a multiplicative coefficient $C = r_g/\nu_0$ in front of the affine parameter in Eq. (\ref{eq_dI}),
\begin{equation}
\frac{d}{d\omega} \rightarrow \frac{1}{C} \frac{d}{d\omega},
\end{equation}
where $r_g = GM$ is the unit length, and $\nu_0$ is the frequency of the real photon at infinity. Eq.~\eqref{eq_dI} then becomes
\begin{equation}
\frac{1}{C} \frac{d}{d\omega}\mathcal{I} =\mathcal{J} - \alpha \mathcal{I},
\end{equation}
and its solution is
\begin{equation}
I_\nu =  r_g \int_{\omega_0}^{\omega} d\omega' g^2 j_\nu(\omega') \exp\big(-r_g \int_{\omega'}^{\omega} d\omega'' \alpha_\nu(\omega'')/g \big)+g^3 I_{\nu_0}\,,\label{eq_inu}
\end{equation}
where $g = \nu_0 / \nu$ is the redshift factor. Assume that the local magnetic field $B^\mu$ follows $B_\mu u^\mu = 0$, the redshift factor becomes
\begin{equation}
g = \frac{k_\alpha (\partial_t)^\alpha}{k_\beta u^\beta} = \frac{k_t}{k_\alpha u^\alpha} = \frac{-1}{k_\alpha u^\alpha}.
\end{equation}
where $k_\mu$ denotes the four-momentum of the reference photon with $k_t=-1$. As is evident from the preceding analysis, evaluation of the intensity requires determination of both the emissivity and the absorption coefficient.
The radiative coefficients $j_\nu$ and $\alpha_\nu$ in Eq.~\eqref{eq_I_J_alpha} are determined by the relevant radiation mechanisms. 

\subsubsection{Synchrotron radiation}
In Eq.~\eqref{eq_I_J_alpha}, the emission coefficient $j_\nu$ and absorption coefficient $\alpha_\nu$ depend explicitly on the underlying radiative processes. In this work, we consider synchrotron emission from electrons in the ultra-relativistic regime, formulated in the Gaussian CGS unit system. Throughout this section, $e$ represents the elementary charge and $k_{\mathrm{B}}$ is the Boltzmann constant.

In a plasma system, synchrotron radiation mainly arises from the contribution of electrons, and its emissivity is given by
\begin{equation}
j_\nu = \frac{\sqrt{3}e^3 B \sin\theta_B}{4\pi m_e} \int_0^\infty d\tau N(\tau) F\left( \nu/\nu_s \right),
\end{equation}
where $\tau = 1 / \sqrt{1 - \beta^2}$ is the Lorentz factor of the charged particle, $N(\tau)$ is the electron distribution function, and $F(x)$ is related to the modified Bessel function of the second kind $K_n(x)$ of order $n$,
\begin{equation}
F(x) = x \int_x^{\infty} dy  K_{5/3}(y).
\end{equation}
Note that the characteristic frequency $\nu_s$ is defined by
\begin{equation}
\nu_s = \frac{3 e B \sin \theta_B \tau^2}{4\pi m c},
\end{equation}
where $\theta_B$ is the angle between $e^\mu_{(B)}$ and $e^\mu_{(k)}$,
\begin{equation}
\theta_B = \arccos \big( e^\mu_{(B)} \cdot e^\mu_{(k)} \big) = \arccos \left[ \frac{g}{B} (B_\mu k^\mu) \right],
\end{equation}
with
\begin{equation}
	e^\mu_{(k)} = -\Big( \frac{k^\mu}{u^\nu k} + u^\mu \Big)~~~\text{and}~~~e^\mu_{(B)} = \frac{B^\mu}{B}.
\end{equation}
In general, different electron distributions correspond to different radiation formulae. In this work, we consider a thermal distribution, whose distribution function is given by
\begin{equation}
N(\tau) = n_e \frac{\tau^2 \beta}{\theta_e K_2(1/\theta_e)} \exp\left( -\tau/\theta_e \right),
\end{equation}
where $n_e$ is the electron number density, $\theta_e = k_B T_e / m_e $ is the dimensionless electron temperature, and $T_e$ is the electron thermodynamic temperature.
For the ultrarelativistic limit, we have $\beta \approx 1$, $\theta_e \gg 1$, the asymptotic formula is $K_2(1/\theta_e) \approx 2\theta_e^2$. Let $y = \tau/\theta_e$ and $s = (\nu/\nu_s)y^{2}$, the emissivity becomes
\begin{equation}
j_\nu = \frac{\sqrt{3}n_e e^3 B \sin\theta_B}{8\pi m_e } \int_0^\infty dy ~y^2 \exp(-y)  F\left( \nu/\nu_s \right)
=\frac{n_e e^2 \nu}{2\sqrt{3}\,\theta_e^2}  I(s),\label{eq_jnu}
\end{equation}
where the characteristic frequency $\nu_c$ of the system is given by
\begin{equation}
\nu_c = \frac{3e B \sin\theta_B \theta_e^2}{4\pi m_e c},
\end{equation}
and the dimensionless function is defined as
\begin{equation}
I(s) = \frac{1}{s} \int_0^\infty y^2 \exp(-y) F\left( s/y^2 \right).
\end{equation}
This equation has no specific analytical formula and requires the use of fitting functions. For isotropic radiation~\cite{leung2011numerical}, the fitting function follows
\begin{equation}
I(s) = 4.0505s^{-1/6} ( 1 + 0.4s^{-1/4} + 0.5316s^{-1/2} ) \times\exp( -1.8899s^{1/3} ),\label{eqI1}
\end{equation}
and for anisotropic radiation~\cite{mahadevan1996harmony}, the fitting function is
\begin{equation}
I(s) = 2.5651 ( 1 + 1.92s^{-1/3} + 0.9977s^{-2/3} )  \times\exp( -1.8899s^{1/3} ).\label{eqI2}
\end{equation}
Against the background of a thermal electron distribution, the absorption process obeys Kirchhoff's law, which states that all absorption coefficients must satisfy
\begin{equation}
\alpha_{\nu} = \frac{j_{\nu}}{B_{\nu}}\quad\text{and} \quad B_{\nu} = \frac{2h\nu^{3}}{\exp(h\nu/k_{B}T_{e}) - 1},\label{eq_alphanu_bnu}
\end{equation}
where \( B_{\nu} \) represents the Planck black body radiation function.
To facilitate the numerical calculations, we introduce the following characteristic constants in the simulation:
\begin{equation}
C_{1}=\frac{n_{h}e^{2}\nu_{h}}{2\sqrt{3}\,\theta_{h}^{2}},\quad
C_{2}=\frac{4\pi m_{e}\nu_{h}}{3eB_{h}\theta_{h}^{2}},\quad
C_{3}=\frac{h\nu_{h}}{m_{e}}\frac{1}{\theta_{h}},\quad
C_{4}=2h\nu_{h}^{3},\quad
C_{5}=\sqrt{n_{h}m_{p}}\,,
\end{equation}
where $n_{\mathrm{h}}$ and $\theta_{\mathrm{h}}$ are evaluated at the event horizon, with characteristic scaling parameters chosen as $\nu_{\mathrm{h}} = 10^9\,\text{Hz} = 1\,\text{GHz}$, and $B_{\mathrm{h}} = 1$. Under this parametrization,
\begin{equation}\label{eq_jxb}
j_{\nu}=C_{1}\bar{n}_{e}\tilde{\nu}\tilde{\theta}_{e}^{-2}I(x),\quad
x=\frac{C_{2}\tilde{\nu}}{\tilde{B}\sin\theta_{B}\tilde{\theta}_{e}^{2}},\quad
B_{\nu}=\frac{C_{4}\tilde{\nu}^{3}}{\exp(C_{3}\tilde{\nu}/\tilde{\theta}_{e})-1},
\end{equation}
where $\bar{\nu} = \nu / \nu_{\mathrm{h}}$, $\bar{n}_e = n_e / n_{\mathrm{h}}$, and $\bar{\theta}_e = \theta_e / \theta_{\mathrm{h}}$ denote the dimensionless normalized quantities, respectively. 

Based on Eqs.~\eqref{eq_alphanu_bnu} and \eqref{eq_jxb}, one can in principle evaluate the specific intensity in Eq.~\eqref{eq_inu}. It should be emphasized, however, that the radial profiles of the electron number density, electron temperature, and the dimensionless synchrotron function $\mathcal{I}(x)$ remain to be explicitly specified. In what follows, we discuss the detailed prescriptions for determining these physical quantities across different accretion disk models.

\subsection{Accretion disk models}\label{sec:4}
For the unpolarized radiative transfer analysis, we adopt two geometrically thick and optically thin accretion flow models: the phenomenological model\cite{yuan2003nonthermal,broderick2011evidence} and the Hou disk model\cite{Hou2024,Zhang2024}.

\subsubsection{Phenomenological model}
The phenomenological model was further developed and refined by Broderick {\em et al.}~\cite{broderick2011evidence} on the basis of the inefficient accretion flow model proposed by Yuan {\em et al.}~\cite{yuan2003nonthermal}.
In this model, the number density and temperature distribution of non-thermal electrons can be respectively described as
\begin{equation}
n_e = n_h \left( r/r_h \right)^2 \exp\left[ -z^2(\alpha R)^{-2}/2 \right]\quad\text{and}\quad
T_e = T_h \left( r/r_h \right),
\label{eq:profiles}
\end{equation}
where \(R = r\sin\theta\) and \(z = r\cos\theta\) denote the cylindrical radius and the vertical height measured from the equatorial plane (\(\theta=\pi/2\)), respectively. Here, \(n_{\mathrm{h}}\), \(\alpha\), and \(T_{\mathrm{h}}\) are constant model-dependent parameters, while \(r_{\mathrm{h}}\) denotes the outer event horizon radius. The magnetic field strength is parameterized by the cold magnetization parameter \(\sigma_{\mathrm{m}}\),
\begin{equation}
\sigma = B^2/\rho,
\end{equation}
with $ \rho = n_e m_p$ the density of the fluid mass. 
The parameter $\sigma$ is a constant, and for the accretion disk model considered here, its magnitude is on the order of  $\sigma \sim 0.1$. 

For the accretion flow motion model, this paper considers the infalling motion~\cite{pu2016effects}. 
Assuming that the fluid is at rest at infinity, i.e., $u_t = -1$, the four-velocity is given by
\begin{equation}
u^\mu = \big( -g^{tt}, -\sqrt{-(1+g^{tt})g^{rr}}\,, 0, 0 \big).
\end{equation}

\subsubsection{Hou disk model}
The Hou disk model is a stationary, axisymmetric accretion flow model proposed in Refs.~\cite{Zhang2024,Hou2024}. Within this framework, the accreting plasma is assumed to be confined to surfaces of constant polar angle $\theta$, which naturally implies that the poloidal four-velocity component vanishes, i.e., $u^\theta \equiv 0$. Consequently, the relativistic mass conservation equation takes the form of
\begin{equation}
\frac{d}{dr} \left( \sqrt{-g} \rho u^r \right) = 0,
\end{equation}
and its solution is
\begin{equation}
\rho = \frac{\rho_0}{\sqrt{-g} u^r}.
\end{equation}
where \(\rho_0 = \rho(r_0)\) is the mass density at the reference point, typically taken at the horizon, \(r_0=r_h\). Projecting the conservation equation of the energy-momentum tensor along \(u^\mu\), we obtain 
\begin{equation}
de = \frac{e + p}{\rho} d\rho,\label{eq_de}
\end{equation}
where $e$ is the internal energy of the fluid. Defining $\kappa= T_p/T_e$ as the proton-to-electron temperature ratio, the internal energy of the fluid under this approximation satisfies the relation
\begin{equation}
e = \rho + \rho \frac{3}{2} (\kappa + 2) \frac{m_e}{m_p} \theta_e,\label{eq_e}
\end{equation}
where $\theta_e = k_B T_e / m_e$ is the dimensionless electron temperature. Using the ideal gas equation of state, we obtain
\begin{equation}
p = n k_B (T_p + T_e) = \rho (1 + \kappa) \frac{m_e}{m_p} \theta_e.\label{eq_p}
\end{equation}
Substituting Eqs.~\eqref{eq_e} and \eqref{eq_p} into Eq.~\eqref{eq_de}, we have
\begin{equation}
\theta_e = (\theta_e)_0 \left( \rho/\rho_0 \right)^{\frac{2(1+\kappa)}{3(2+\kappa)}},
\end{equation}
where $(\theta_e)_0 = \theta_e(r_0)$ is the temperature at the reference point, again chosen as $r_0 = r_h$.
For convenience, we assume that $\rho(r_h, \theta)$ follows a Gaussian distribution in the $\theta$ direction, and set $\theta_e(r_h, \theta)$ as a constant in the conical solution,
\begin{equation}
\rho(r_h,\theta) = \rho_h \exp\left[-\left(\frac{\sin\theta-\sin\theta_J}{\sigma}\right)^2\right]\quad\text{and}\quad
\theta(r_h,\theta) = \theta_h,
\end{equation}
where $\theta_J$ is the mean position in the $\theta$ direction, and $\sigma$ describes the standard deviation of the distribution. Given the mass density of the mass, the number density can be obtained from the relation $\rho = n_e m_p$.
For M$\mathrm{87}^*$, observations indicate that $\rho_h \approx 1.5 \times 10^3  \,\mathrm{g/cm/s^2}$ with $\theta_h \approx 16.86$, $n_h = 10^6  \,\mathrm{cm^{-3}}$, and $T_h = 10^{11}  \,\mathrm{K}$. 

The general configuration of the magnetic field can be derived under the steady-state axisymmetric condition,
\begin{equation}
B^\mu = \frac{\Psi}{\sqrt{-g} u^r} (u_t u^\mu + \delta^\mu_t),
\end{equation}
where \(\Psi= F_{\theta\phi}\) is a component of the electromagnetic tensor. Note that \(u^r\) appears in the denominator. Therefore, orbital motion is not permitted for the fluid in the Hou disk model. Here, we adopt a split-monopole solution,
\begin{equation}
\Psi = \Psi_0\, \mathrm{sign}(\cos \theta) \sin \theta.
\end{equation}
It indicates that magnetic-field orientation must be taken into account for the Hou disk model. Accordingly, the anisotropic radiation formula~\eqref{eqI2} is adopted for the electron radiation model.

\section{Non-polarized imaging}\label{sec:4}
In the non-polarized case, two geometrically thick but optically thin accretion disk models, the phenomenological model and the Hou disk model, are considered in this section.

\subsection{Phenomenological model}
It is worth emphasizing that synchrotron radiation is intrinsically anisotropic, and its emissivity depends strongly on the emission direction. This dependence is governed by the pitch angle \(\theta_B\), defined as the angle between the photon wave vector and the magnetic field in the fluid rest frame. To quantitatively assess the impact of the anisotropy on black hole images, we first perform a comparative analysis with an isotropic emission model, which serves as a baseline reference to isolate and evaluate the effects induced by anisotropy.

\begin{figure}[!htb]
  \centering
  \begin{subfigure}{0.29\textwidth}
    \includegraphics[width=\textwidth,height=0.3\textheight,keepaspectratio]{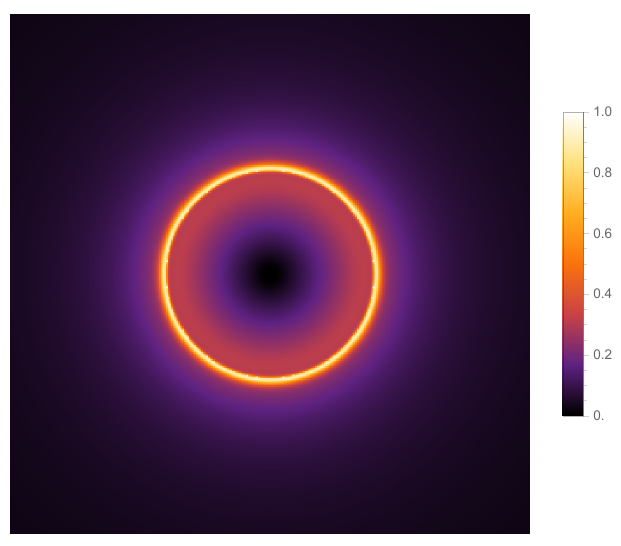}
    \caption{$\alpha=0.01$, $\theta=0.1^\circ$}
  \end{subfigure}
  \quad\quad
  \begin{subfigure}{0.29\textwidth}
    \includegraphics[width=\textwidth,height=0.3\textheight,keepaspectratio]{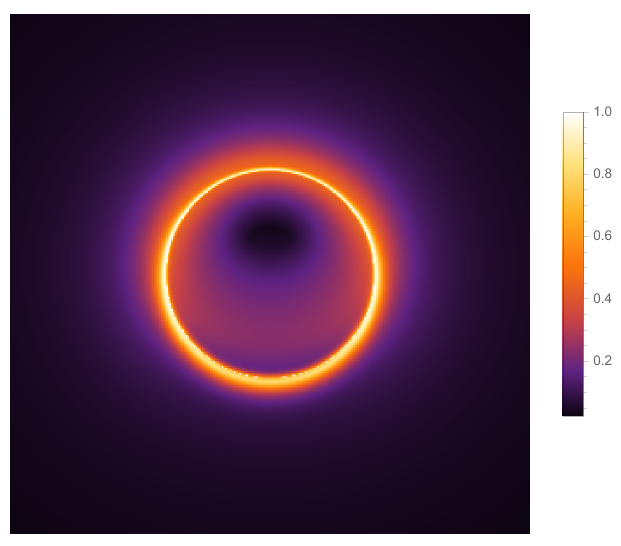}
    \caption{$\alpha=0.01$, $\theta=45^\circ$}
  \end{subfigure}
  \quad\quad
  \begin{subfigure}{0.29\textwidth}
    \includegraphics[width=\textwidth,height=0.3\textheight,keepaspectratio]{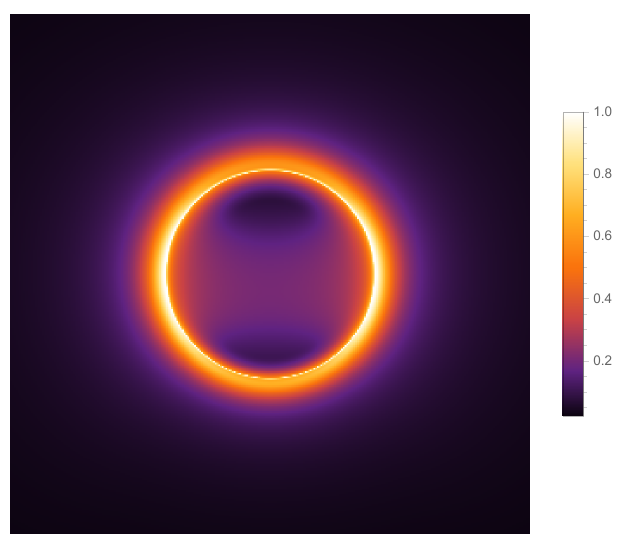}
    \caption{$\alpha=0.01$, $\theta=80^\circ$}
  \end{subfigure}
  \begin{subfigure}{0.29\textwidth}
    \includegraphics[width=\textwidth,height=0.3\textheight,keepaspectratio]{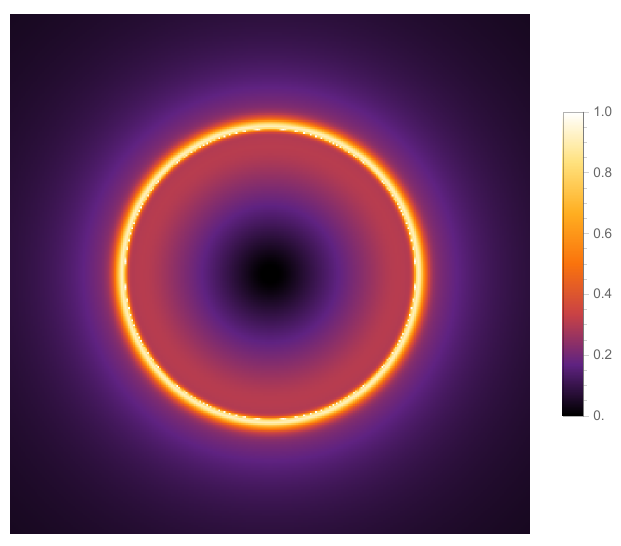}
    \caption{$\alpha=0.5$, $\theta=0.1^\circ$}
  \end{subfigure}
  \quad\quad
 \begin{subfigure}{0.29\textwidth}
    \includegraphics[width=\textwidth,height=0.3\textheight,keepaspectratio]{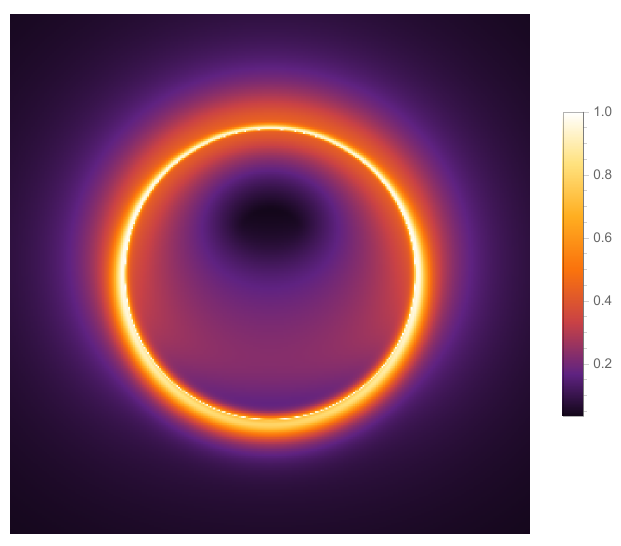}
    \caption{$\alpha=0.5$, $\theta=45^\circ$}
  \end{subfigure}
  \quad\quad
  \begin{subfigure}{0.29\textwidth}
    \includegraphics[width=\textwidth,height=0.3\textheight,keepaspectratio]{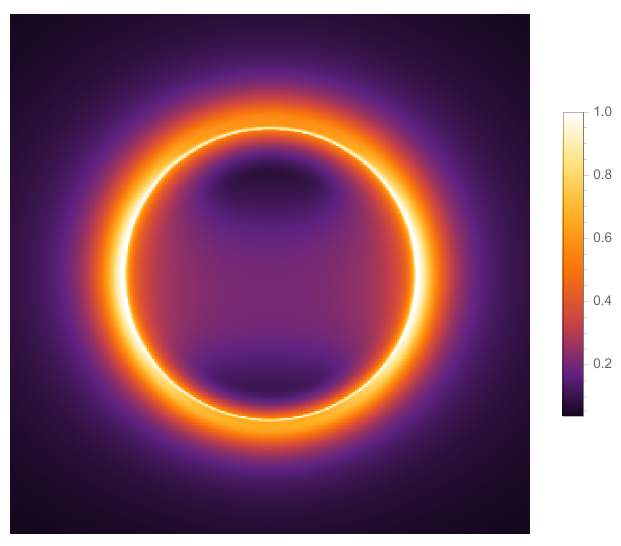}
    \caption{$\alpha=0.5$, $\theta=80^\circ$}
  \end{subfigure}
  \begin{subfigure}{0.29\textwidth}
    \includegraphics[width=\textwidth,height=0.3\textheight,keepaspectratio]{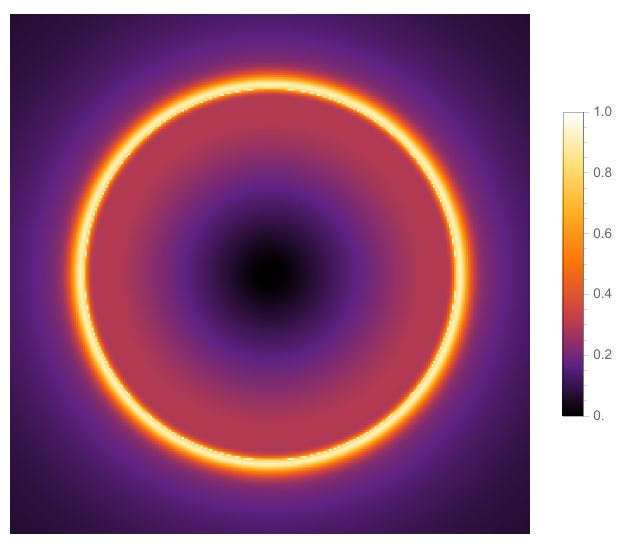}
    \caption{$\alpha=0.99$, $\theta=0.1^\circ$}
  \end{subfigure}
  \quad\quad
  \begin{subfigure}{0.29\textwidth}
    \includegraphics[width=\textwidth,height=0.3\textheight,keepaspectratio]{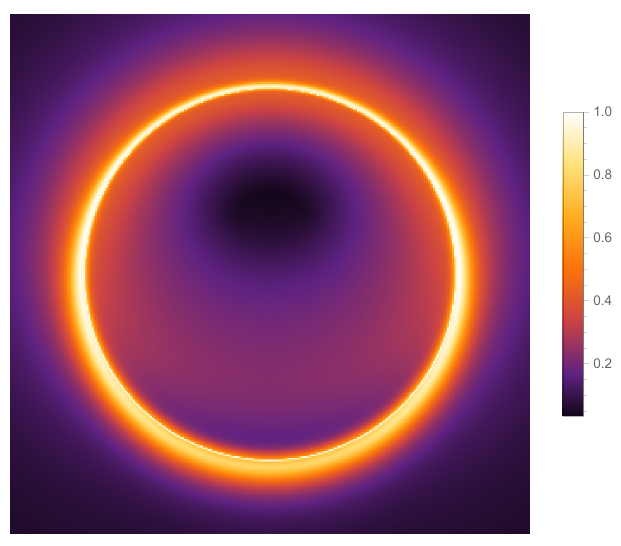}
    \caption{$\alpha=0.99$, $\theta=45^\circ$}
  \end{subfigure}
  \quad\quad
  \begin{subfigure}{0.29\textwidth}
    \includegraphics[width=\textwidth,height=0.3\textheight,keepaspectratio]{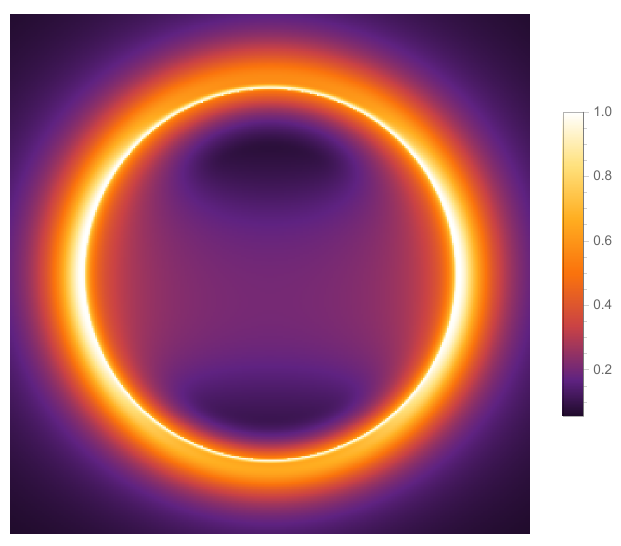}
    \caption{$\alpha=0.99$, $\theta=80^\circ$}
  \end{subfigure}
   \caption{Simulated shadow images and intensity distributions of the black hole within the phenomenological model under isotropic radiation and an infalling accretion flow. The observer distance is fixed at $r_{\mathrm{obs}} = 500M$, the field of view is $3^\circ$, and the observation frequency is $230\,\text{GHz}$.}
   \label{figtong}
\end{figure}
As illustrated in Fig.~\ref{figtong}, a prominent bright ring appears in all simulated images, corresponding to higher-order lensing features formed by photons that complete one or more half-loops around the black hole before reaching the observer. Outside this ring, the extended region with non-zero intensity represents direct (primary) emission, originating from photons traveling straight from the accretion flow to the observer. 

Notably, across all parameter configurations, a region of vanishing intensity emerges interior to the higher-order ring, cast by the black hole event horizon. For a geometrically thin accretion disk, this feature corresponds to the canonical ``inner shadow'', which provides a potential observational signature for the EHT\cite{akiyama2019firstm87,akiyama2019first5,akiyama2024event}. However, for the geometrically thick accretion flows considered in this work, this region can be partially filled or obscured by off-equatorial emission, preventing its unambiguous identification. Given that thick-disk geometries generally offer a more physically realistic description of astrophysical accretion systems, our results highlight that directly imaging the silhouette of the event horizon remains observationally challenging.

\begin{figure}[!htb]
  \centering
  \begin{subfigure}{0.29\textwidth}
    \includegraphics[width=\textwidth,height=0.3\textheight,keepaspectratio]{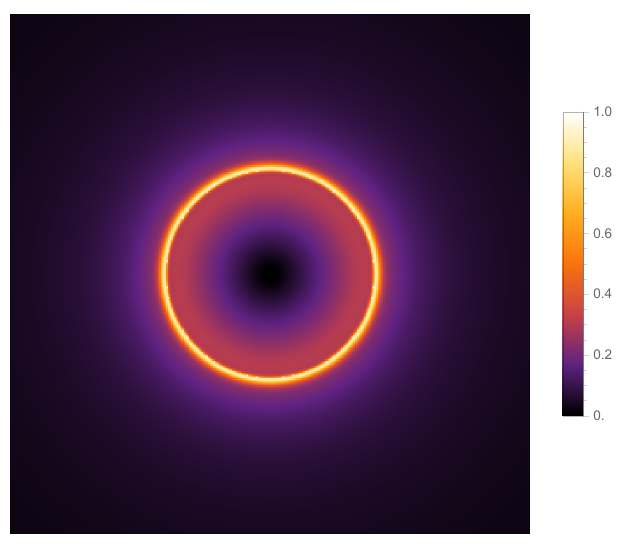}
    \caption{$\alpha=0.01$, $\theta=0.1^\circ$}
  \end{subfigure}
  \quad\quad
  \begin{subfigure}{0.29\textwidth}
    \includegraphics[width=\textwidth,height=0.3\textheight,keepaspectratio]{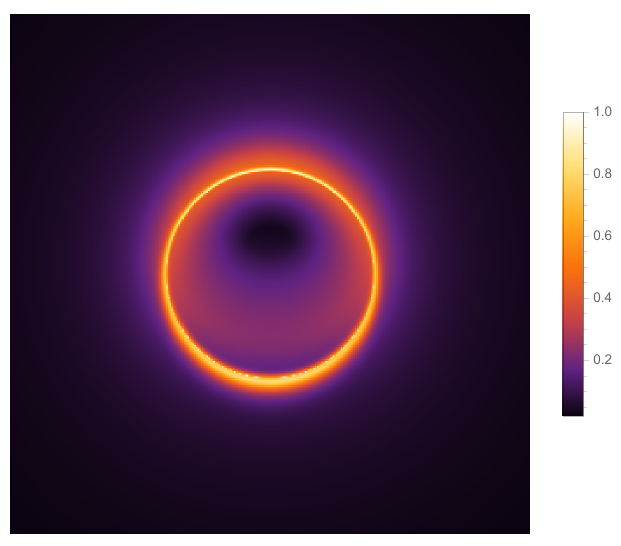}
    \caption{$\alpha=0.01$, $\theta=45^\circ$}
  \end{subfigure}
  \quad\quad
  \begin{subfigure}{0.29\textwidth}
    \includegraphics[width=\textwidth,height=0.3\textheight,keepaspectratio]{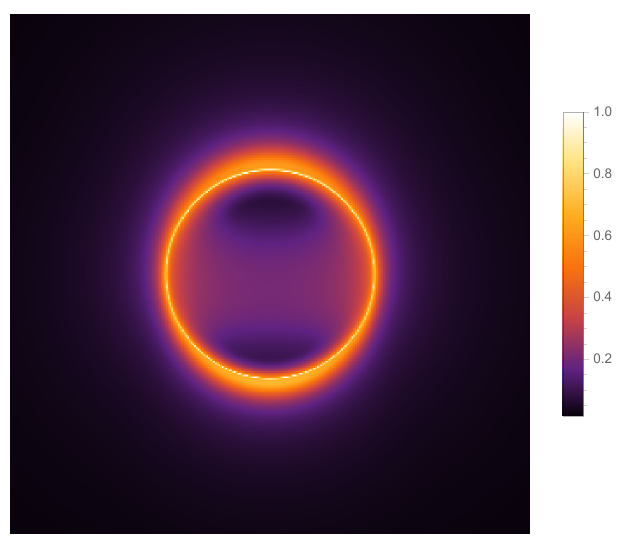}
    \caption{$\alpha=0.01$, $\theta=80^\circ$}
  \end{subfigure}
  \begin{subfigure}{0.29\textwidth}
    \includegraphics[width=\textwidth,height=0.3\textheight,keepaspectratio]{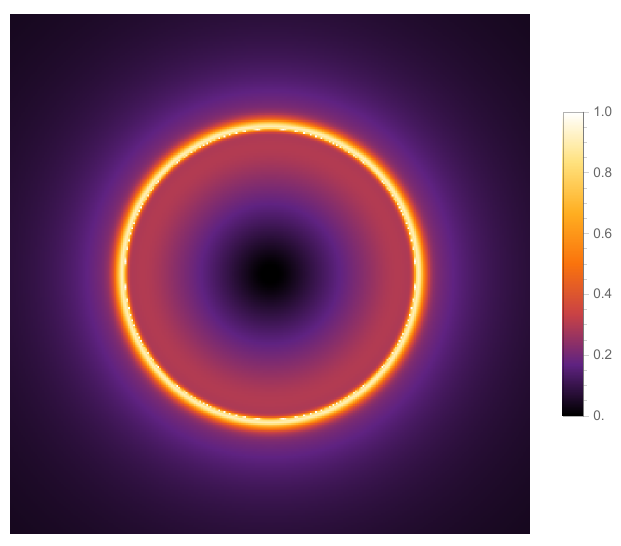}
    \caption{$\alpha=0.5$, $\theta=0.1^\circ$}
  \end{subfigure}
  \quad\quad
 \begin{subfigure}{0.29\textwidth}
    \includegraphics[width=\textwidth,height=0.3\textheight,keepaspectratio]{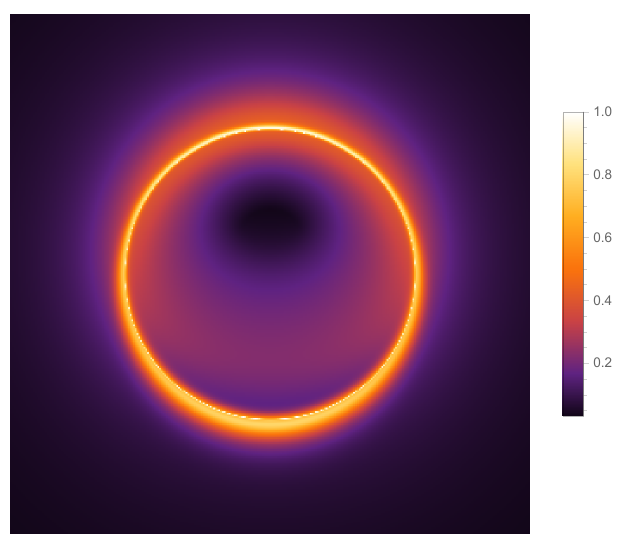}
    \caption{$\alpha=0.5$, $\theta=45^\circ$}
  \end{subfigure}
  \quad\quad
  \begin{subfigure}{0.29\textwidth}
    \includegraphics[width=\textwidth,height=0.3\textheight,keepaspectratio]{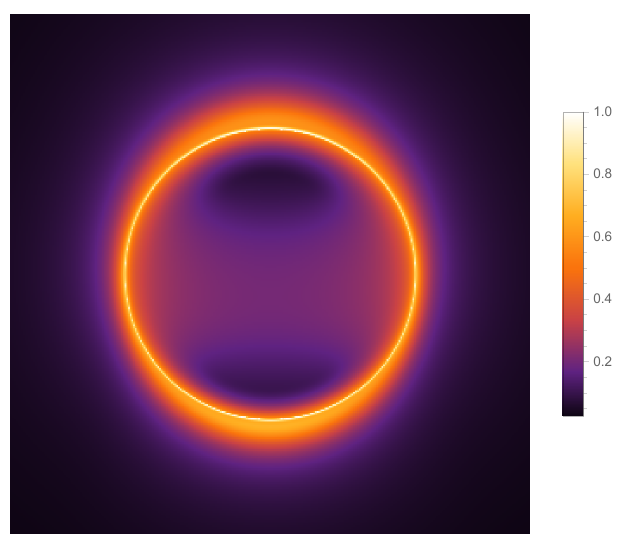}
    \caption{$\alpha=0.5$, $\theta=80^\circ$}
  \end{subfigure}
  \begin{subfigure}{0.29\textwidth}
    \includegraphics[width=\textwidth,height=0.3\textheight,keepaspectratio]{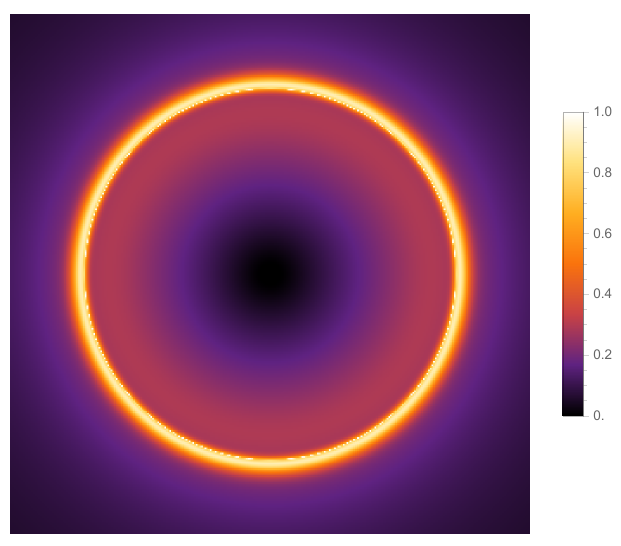}
    \caption{$\alpha=0.99$, $\theta=0.1^\circ$}
  \end{subfigure}
  \quad\quad
  \begin{subfigure}{0.29\textwidth}
    \includegraphics[width=\textwidth,height=0.3\textheight,keepaspectratio]{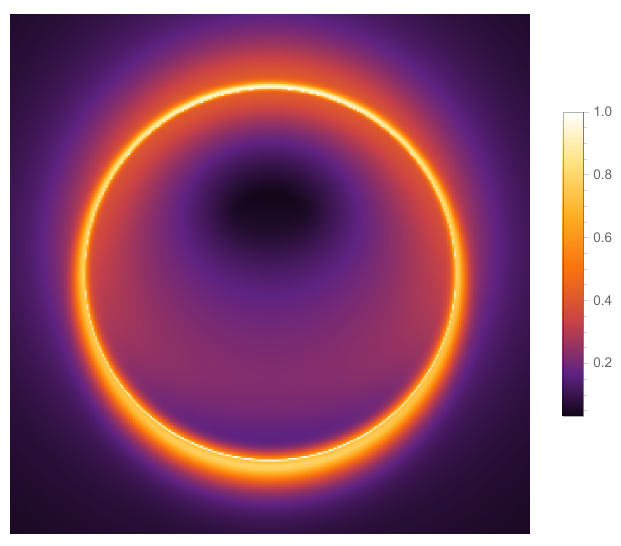}
    \caption{$\alpha=0.99$, $\theta=45^\circ$}
  \end{subfigure}
  \quad\quad
  \begin{subfigure}{0.29\textwidth}
    \includegraphics[width=\textwidth,height=0.3\textheight,keepaspectratio]{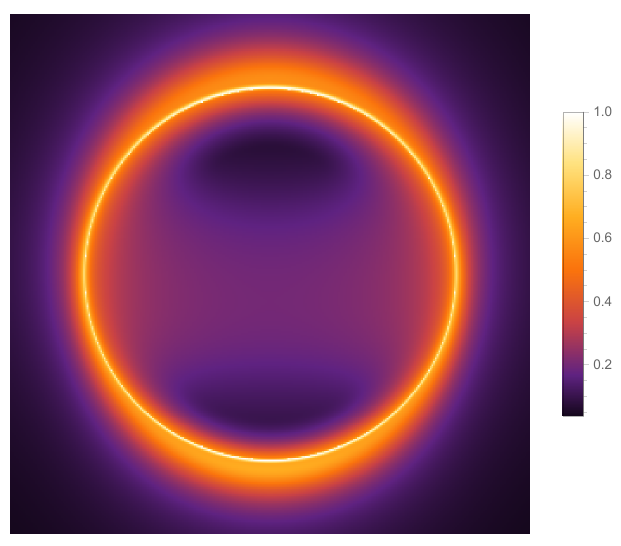}
    \caption{$\alpha=0.99$, $\theta=80^\circ$}
  \end{subfigure}
   \caption{Black hole shadow images for the phenomenological model under anisotropic radiation. The observer distance is fixed at 500$M$, the field of view is $3^\circ$, and the observation frequency is $230\,\text{GHz}$.}
   \label{figyi}
\end{figure}
We next extend our analysis to incorporate anisotropic synchrotron radiation. As shown in Fig.~\ref{figyi}, the background spacetime parameters affect the black hole shadow under anisotropic emission in a manner broadly consistent with the isotropic case. As the viewing inclination angle $\theta$ increases, both the apparent size and flux intensity of the higher-order lensed images gradually decrease. Concurrently, the morphology of the higher-order ring undergoes notable structural distortion with increasing $\theta$, where off-equatorial emission increasingly obscures the central dark region interior to the photon ring.

\subsection{Hou disk model}
This model assumes that fluid acceleration in the near-horizon regime is predominantly governed by gravity. Given the specified magnetic field configuration, electron number density, and temperature distributions, one can in principle integrate the radiative transfer equation~\eqref{eq_inu} numerically using the emissivity and absorption prescriptions given in Eqs.~\eqref{eq_jnu} and \eqref{eq_alphanu_bnu}. In the following, we present the resulting numerical solutions for the Hou disk model.

\begin{figure}[!htb]
  \centering
  \begin{subfigure}{0.29\textwidth}
    \centering
    \includegraphics[width=\textwidth,height=0.3\textheight,keepaspectratio]{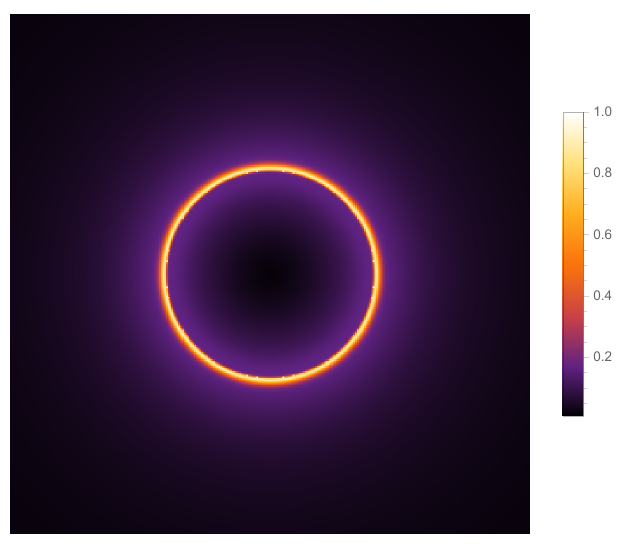}
    \caption{$\alpha=0.01$, $\theta=0.1^\circ$}
  \end{subfigure}
  \quad\quad
  \begin{subfigure}{0.29\textwidth}
    \centering
    \includegraphics[width=\textwidth,height=0.3\textheight,keepaspectratio]{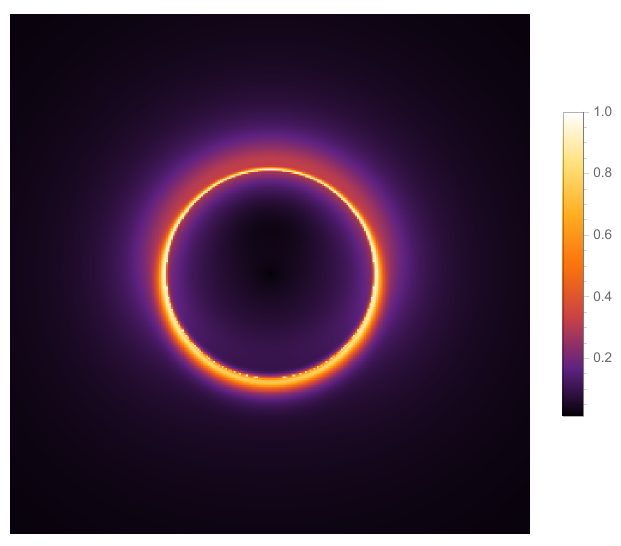}
   \caption{$\alpha=0.01$, $\theta=45^\circ$}
  \end{subfigure}
  \quad\quad
  \begin{subfigure}{0.29\textwidth}
    \centering
    \includegraphics[width=\textwidth,height=0.3\textheight,keepaspectratio]{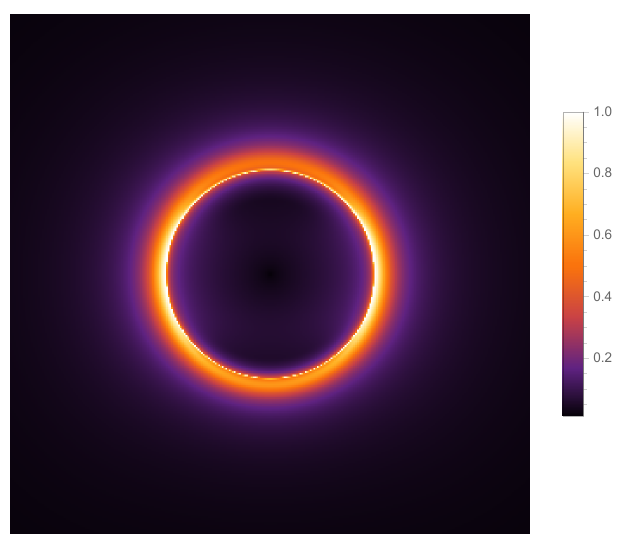}
  \caption{$\alpha=0.01$, $\theta=80^\circ$}
  \end{subfigure}
  \begin{subfigure}{0.29\textwidth}
    \centering
    \includegraphics[width=\textwidth,height=0.3\textheight,keepaspectratio]{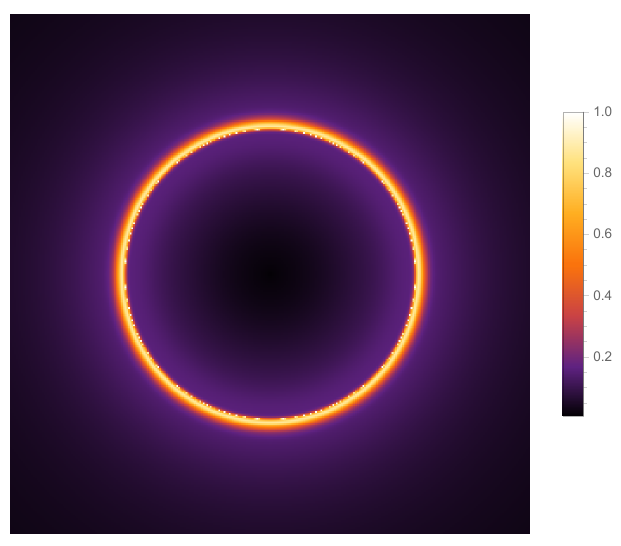}
    \caption{$\alpha=0.5$, $\theta=0.1^\circ$}
  \end{subfigure}
  \quad\quad
  \begin{subfigure}{0.29\textwidth}
    \centering
    \includegraphics[width=\textwidth,height=0.3\textheight,keepaspectratio]{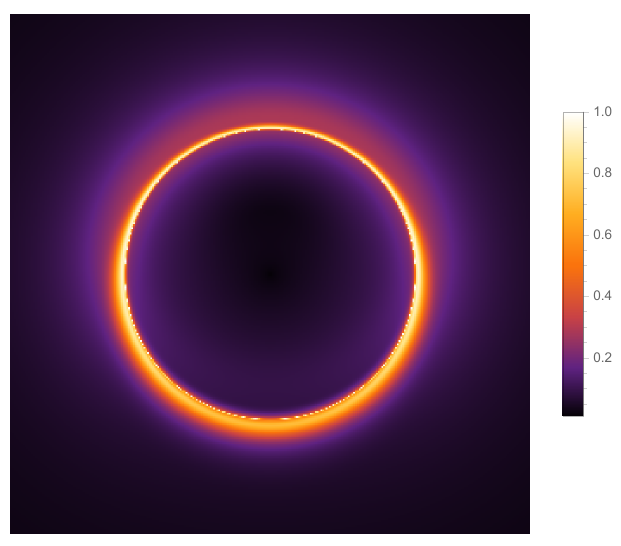}
    \caption{$\alpha=0.5$, $\theta=45^\circ$}
  \end{subfigure}
  \quad\quad
  \begin{subfigure}{0.29\textwidth}
    \centering
    \includegraphics[width=\textwidth,height=0.3\textheight,keepaspectratio]{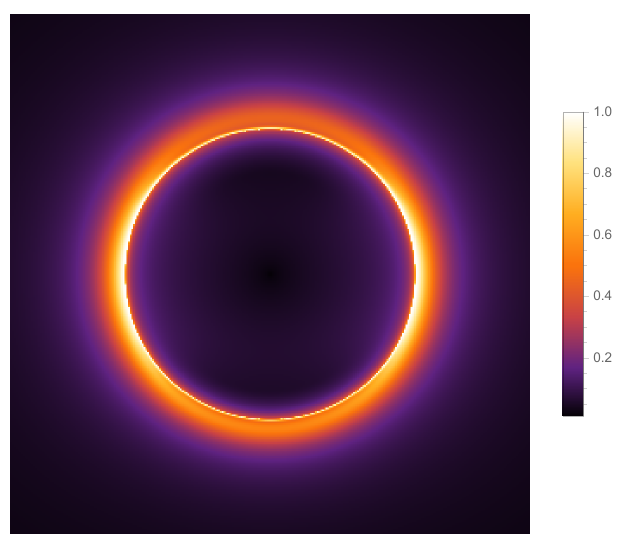}
    \caption{$\alpha=0.5$, $\theta=80^\circ$}
  \end{subfigure}
  \begin{subfigure}{0.29\textwidth}
    \centering
    \includegraphics[width=\textwidth,height=0.3\textheight,keepaspectratio]{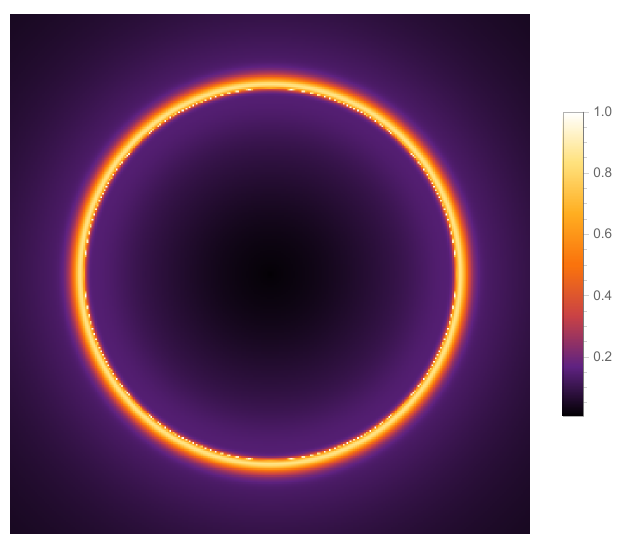}
    \caption{$\alpha=0.99$, $\theta=0.1^\circ$}
  \end{subfigure}
  \quad\quad
  \begin{subfigure}{0.29\textwidth}
    \centering
    \includegraphics[width=\textwidth,height=0.3\textheight,keepaspectratio]{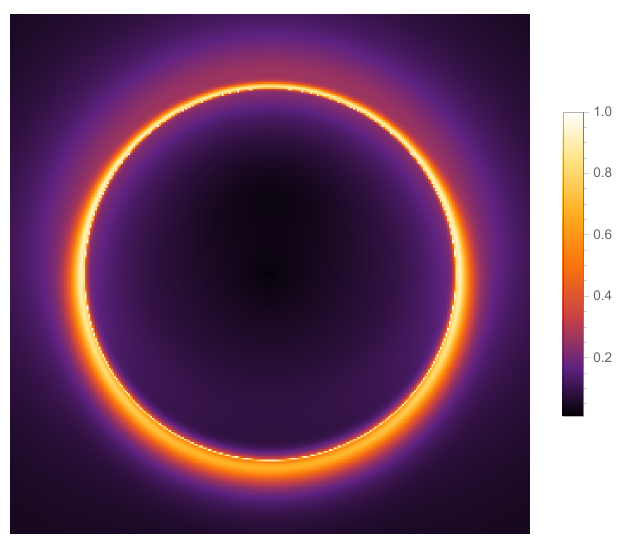}
    \caption{$\alpha=0.99$, $\theta=45^\circ$}
  \end{subfigure}
  \quad\quad
  \begin{subfigure}{0.29\textwidth}
    \centering
    \includegraphics[width=\textwidth,height=0.3\textheight,keepaspectratio]{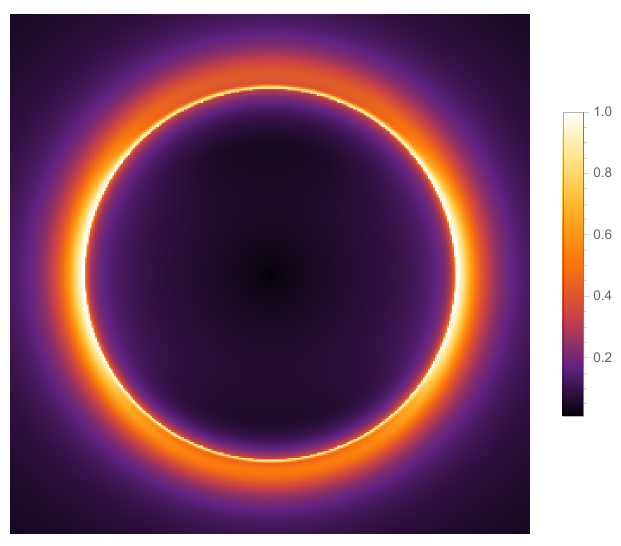}
    \caption{$\alpha=0.99$, $\theta=80^\circ$}
  \end{subfigure}
  \caption{Synthetic shadow images for an infalling accretion flow within the Hou disk model. The observer distance is fixed at $r_{\mathrm{obs}} = 500M$, the field of view is $3^\circ$, and the observing frequency is $ 230\,\text{GHz}$.}
  \label{fig_hou}
\end{figure}

As shown in Fig.~\ref{fig_hou}, a prominent bright ring structure is consistently observed across all panels, corresponding to higher-order lensed images formed by photons executing one or more half-loops around the black hole before reaching the observer. The surrounding extended region with non-zero intensity represents the primary (direct) emission originating from photons traveling directly from the accretion flow to the observer. Notably, interior to the higher-order photon ring, a central dark region with vanishing intensity emerges, cast by the black hole event horizon. In geometrically thin-disk configurations, this feature corresponds to the canonical ``inner shadow''. However, for the geometrically thick accretion flows considered here, this central shadow can be partially obscured and filled by off-equatorial radiation, impeding its clear identification. Given that thick accretion disks provide a more realistic description of low-luminosity astrophysical systems, this result highlights that directly resolving the image of the event horizon remains observationally challenging.

Fig.~\ref{fig_hou} also indicates that as $\alpha$ increases (from top to bottom), both the higher-order ring and the central dark shadow expand in size. Meanwhile, increasing the observer inclination angle $\theta$ (from left to right) systematically enhances the overall flux intensity of the higher-order images. Specifically, at a near face-on inclination of $\theta = 0.1^\circ$ (first column), the higher-order image exhibits a nearly symmetric circular ring. As $\theta$ increases to $45^\circ$ (second column), the lensed ring shifts toward the lower portion of the image plane due to geometric projection and off-equatorial obscuration. At an edge-on viewing angle of $\theta = 80^\circ$ (third column), the intensity along the horizontal (equatorial) direction becomes substantially brighter than that along the vertical axis, a consequence of strong relativistic Doppler beaming along the line of sight. These features demonstrate that the parameter $\alpha$ primarily dictates the spatial scale of the shadow and photon ring, whereas the viewing inclination $\theta$ predominantly modulates the ring morphology and brightness distribution.

\begin{figure}[!htb]
  \centering
    \begin{subfigure}{0.29\textwidth}
    \includegraphics[width=\textwidth,height=0.3\textheight,keepaspectratio]{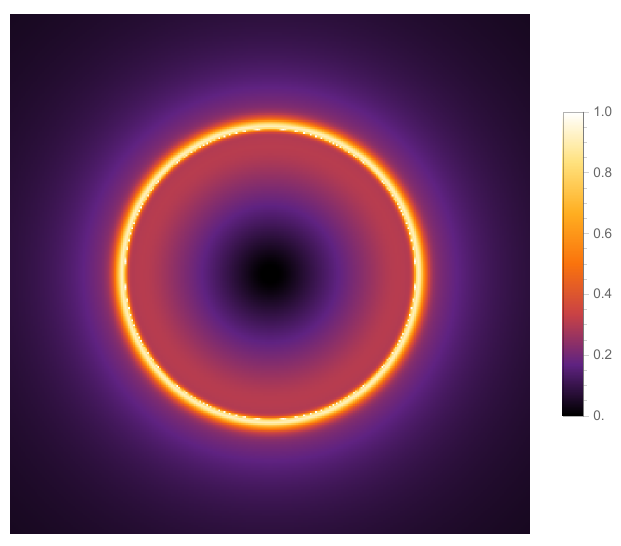}
    \caption{$\theta=0.1^\circ$}
  \end{subfigure}
  \quad\quad
 \begin{subfigure}{0.29\textwidth}
    \includegraphics[width=\textwidth,height=0.3\textheight,keepaspectratio]{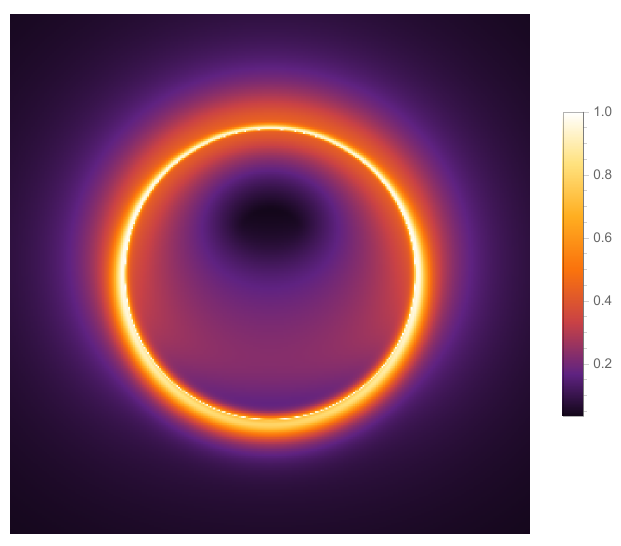}
    \caption{$\theta=45^\circ$}
  \end{subfigure}
  \quad\quad
  \begin{subfigure}{0.29\textwidth}
    \includegraphics[width=\textwidth,height=0.3\textheight,keepaspectratio]{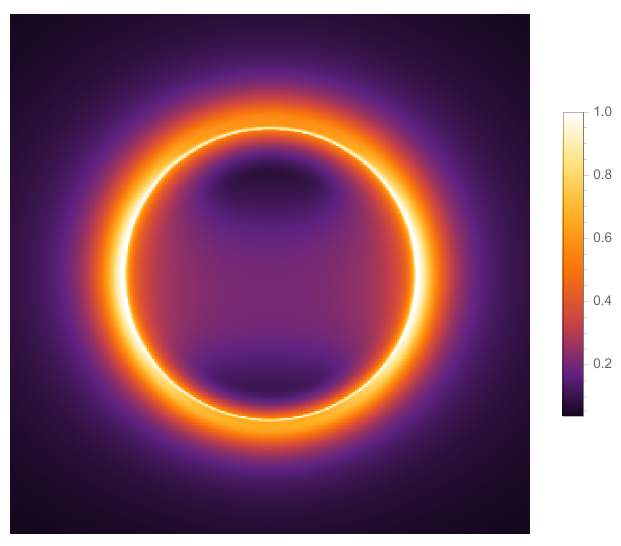}
    \caption{$\theta=80^\circ$}
  \end{subfigure}
   \begin{subfigure}{0.29\textwidth}
    \includegraphics[width=\textwidth,height=0.3\textheight,keepaspectratio]{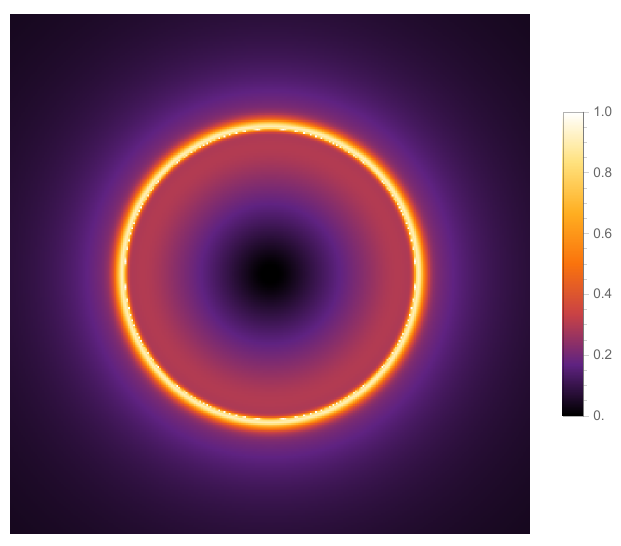}
    \caption{$\theta=0.1^\circ$}
  \end{subfigure}
  \quad\quad
 \begin{subfigure}{0.29\textwidth}
    \includegraphics[width=\textwidth,height=0.3\textheight,keepaspectratio]{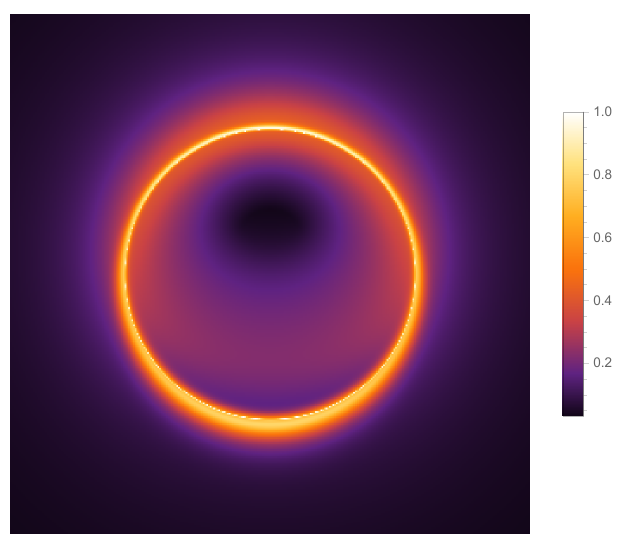}
    \caption{$\theta=45^\circ$}
  \end{subfigure}
  \quad\quad
  \begin{subfigure}{0.29\textwidth}
    \includegraphics[width=\textwidth,height=0.3\textheight,keepaspectratio]{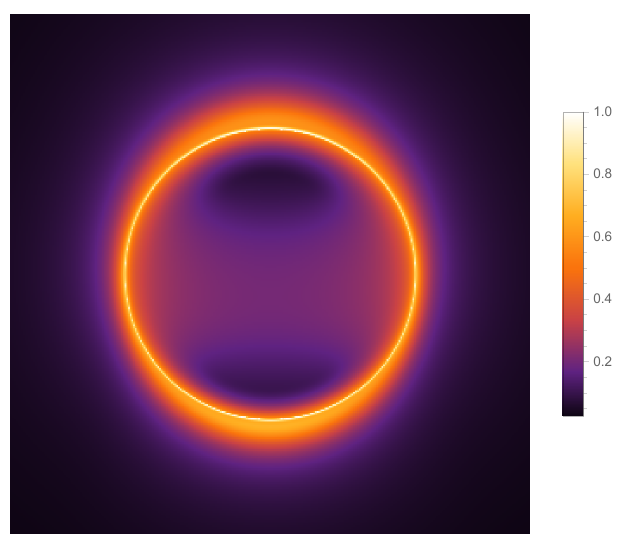}
    \caption{$\theta=80^\circ$}
  \end{subfigure}
  \begin{subfigure}{0.29\textwidth}
    \centering
    \includegraphics[width=\textwidth,height=0.3\textheight,keepaspectratio]{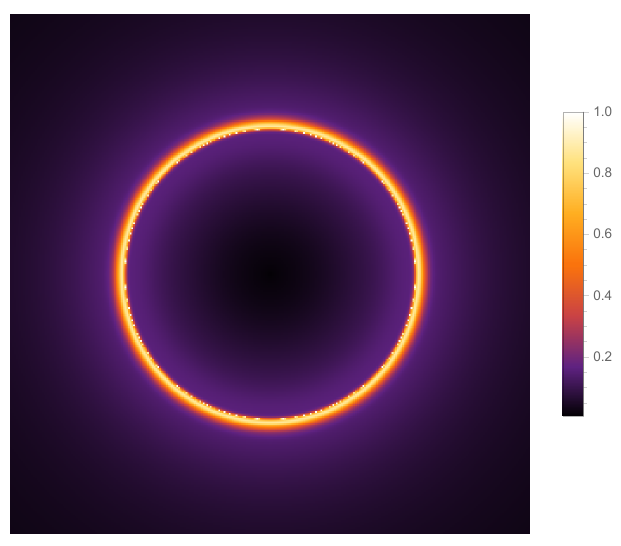}
    \caption{$\theta=0.1^\circ$}
  \end{subfigure}
  \quad\quad
  \begin{subfigure}{0.29\textwidth}
    \centering
    \includegraphics[width=\textwidth,height=0.3\textheight,keepaspectratio]{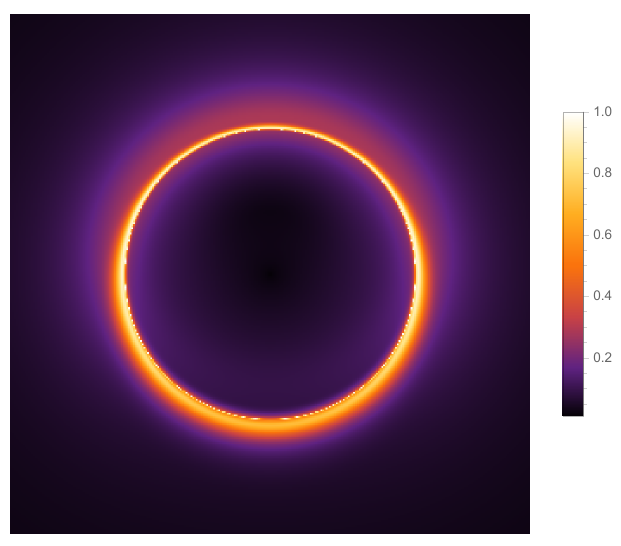}
    \caption{$\theta=45^\circ$}
  \end{subfigure}
  \quad\quad
  \begin{subfigure}{0.29\textwidth}
    \centering
    \includegraphics[width=\textwidth,height=0.3\textheight,keepaspectratio]{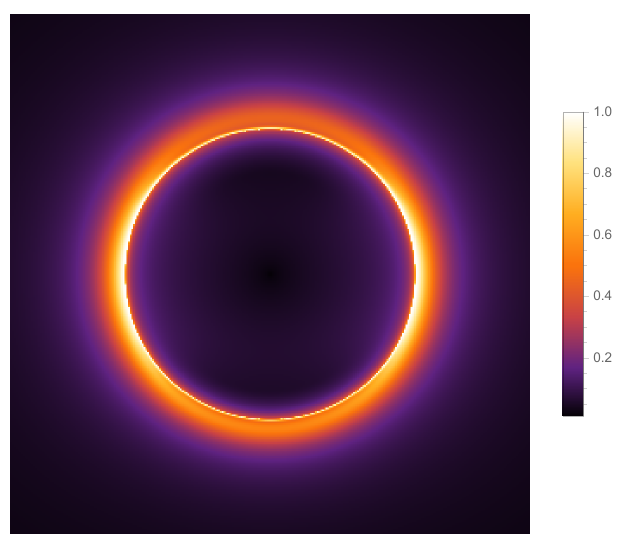}
    \caption{$\theta=80^\circ$}
  \end{subfigure}
  \caption{Influence of accretion disk models on black hole shadows at different observation inclinations. The first and second rows correspond to the phenomenological model with isotropic radiation and anisotropic radiation, respectively. And the third row corresponds to the Hou disk model. The accretion flow is infalling with $\alpha=0.5$ and $\nu=230\,\text{GHz}$.}
  \label{fig_duibi}
\end{figure}
Fig.~\ref{fig_duibi} intuitively compares the different between the three previous models. For a given inclination $\theta$, the phenomenological model with anisotropic emission yields the largest higher-order photon ring. Furthermore, the direct (primary) emission in the phenomenological model exhibits a substantially higher brightness than that in the Hou disk model. Notably, at high inclination angles (e.g., $\theta = 80^\circ$, as shown in the third row), the obscuration of the event horizon silhouette by off-equatorial radiation is significantly more pronounced in the phenomenological model, indicating enhanced gravitational lensing and optical-depth projection effects in this configuration.

\begin{figure}[!htb]
  \centering
  \begin{subfigure}{0.29\textwidth}
    \centering
    \includegraphics[width=\textwidth,height=0.3\textheight,keepaspectratio]{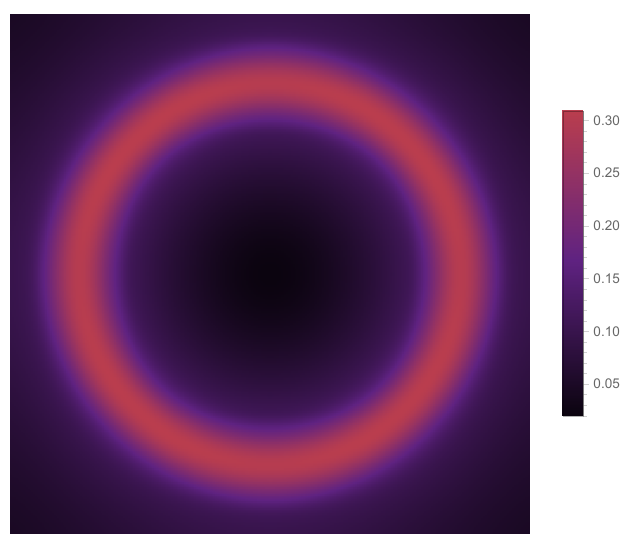}
    \caption{$\alpha=0.99$, $\theta=0.1^\circ$}
  \end{subfigure}
  \quad\quad
  \begin{subfigure}{0.29\textwidth}
    \centering
    \includegraphics[width=\textwidth,height=0.3\textheight,keepaspectratio]{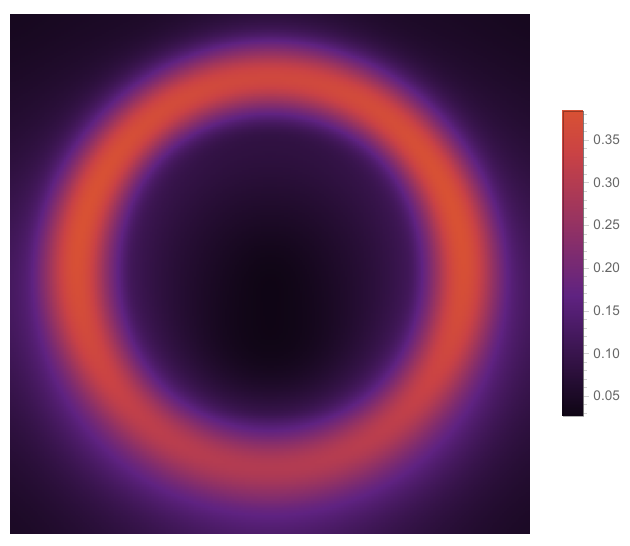}
   \caption{$\alpha=0.99$, $\theta=45^\circ$}
  \end{subfigure}
  \quad\quad
  \begin{subfigure}{0.29\textwidth}
    \centering
    \includegraphics[width=\textwidth,height=0.3\textheight,keepaspectratio]{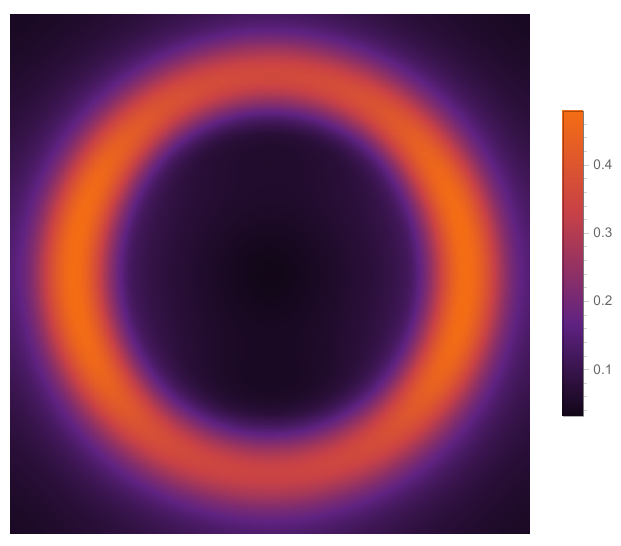}
  \caption{$\alpha=0.99$, $\theta=80^\circ$}
  \end{subfigure}
  \caption{Blurred images of the black hole. From left to right, the observer inclination angles $\theta$ are $0.1^\circ$, $45^\circ$, and $80^\circ$, respectively. The observer distance is fixed at $500M$, the field of view is $3^\circ$, the observing frequency is $230\,\text{GHz}$, and $\alpha=0.99$.}
  \label{fig_houmo}
\end{figure}
Fig.~\ref{fig_houmo} facilitates direct comparison with realistic observations, where the raw simulated images are convolved with a Gaussian beam to account for the finite angular resolution of the Event Horizon Telescope (EHT) for M$87^*$~\cite{akiyama2019firstm87,akiyama2019first5,akiyama2024event}. One can check that the blurring operation washes out the infinitely thin higher-order photon ring and broadens the bright emission region into an extended, diffuse ring morphology, which shows qualitative consistency with the observed horizon-scale features reported by the EHT~\cite{akiyama2019firstm87,akiyama2019first5,akiyama2024event}.

\begin{figure}[!htb]
  \centering
  \begin{subfigure}{0.35\textwidth}
    \centering
    \includegraphics[width=\textwidth,height=0.4\textheight,keepaspectratio]{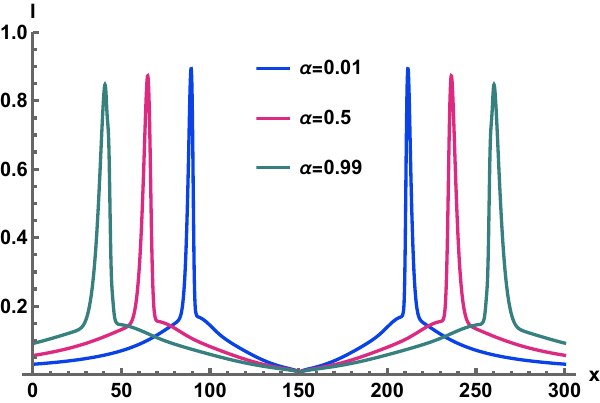}
    \caption{horizontal direction, \(\theta=0.1^\circ\)}
  \end{subfigure}
  \quad\quad
  \begin{subfigure}{0.35\textwidth}
    \centering
    \includegraphics[width=\textwidth,height=0.4\textheight,keepaspectratio]{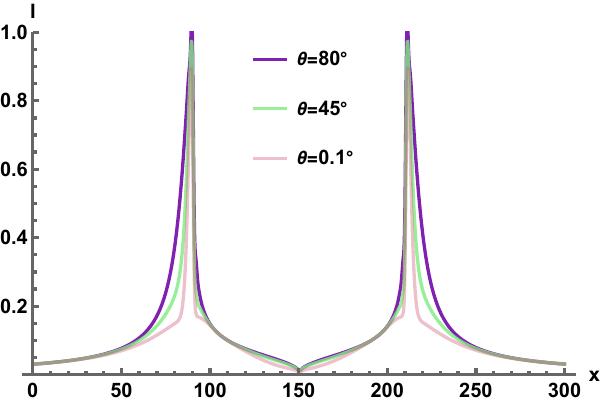}
   \caption{horizontal direction, \(\alpha=0.01\)}
  \end{subfigure}
  \begin{subfigure}{0.36\textwidth}
    \centering
    \includegraphics[width=\textwidth,height=0.4\textheight,keepaspectratio]{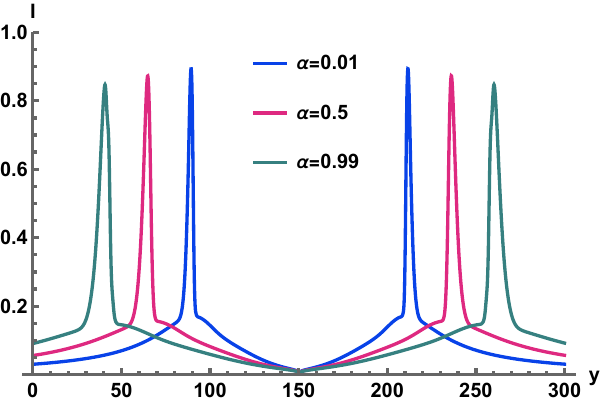}
    \caption{vertical direction, \(\theta=0.1^\circ\)}
  \end{subfigure}
  \quad\quad
  \begin{subfigure}{0.36\textwidth}
    \centering
    \includegraphics[width=\textwidth,height=0.4\textheight,keepaspectratio]{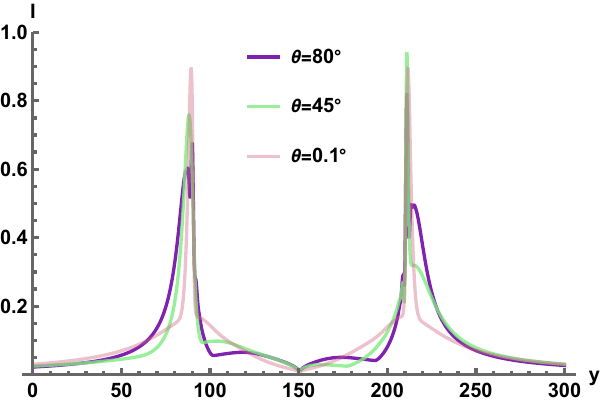}
    \caption{vertical direction, \(\alpha=0.01\)}
  \end{subfigure}
   \caption{Light intensity profiles under different parameter configurations.  The observer distance is fixed at $500M$, and the field of view is set to \(3^\circ\).}
  \label{fig_houcurve}
\end{figure}
As shown in Fig.~\ref{fig_houcurve}, the peak intensity declines and shifts radially outward when \(\alpha\) increases. It indicates that the enhanced  strength of gravitational field will increase the photon capture cross-section and radiative energy absorption, thereby reducing the observed peak flux at infinity. Meanwhile, it will pushes the photon sphere outward, resulting in the radial displacement of the intensity peak. This evolutionary trend is consistent with the results in the first column of Fig.~\ref{fig_hou}, where the expansion of higher-order photon rings is observed with increasing \(\alpha\). Additionally, the horizontal and vertical intensity profiles are nearly identical under the face-on configuration (\(\theta = 0.1^\circ\)), as the system exhibits approximate axial symmetry along the black hole spin axis.

Fig.~\ref{fig_houcurve} also shows that with an increasing inclination angle, both the spatial extension of the emission region and the peak intensity increase in the horizontal direction. Indeed a larger inclination angle requires greater deflection of photons via gravitational lensing, therefore broadens the projected luminous region on the observer's plane. Furthermore, horizontal accretion disk motion generates prominent line-of-sight velocity components: the approaching gas side produces significant relativistic Doppler boosting and brightening, which dominates the overall peak flux and overwhelms the dimming effect of the receding side.

In the vertical direction, the peak intensity is consistently lower than the horizontal counterpart. Increasing \(\theta\) widens the vertical emission range and induces a distinct vertical asymmetry, with the upper-side peak intensity gradually weaker than the lower-side one. For geometrically thick accretion disks, gravitational energy dissipation concentrates high-density and high-temperature plasma near the equatorial plane, forming a radially stratified emission structure with decreasing luminosity toward the polar regions. Accordingly, vertical cross-sections mainly probe the dimmer outer envelope of the disk, leading to intrinsically lower peak intensities. Moreover, pronounced self-obscuration asymmetry emerges at high inclinations: radiation from the hot inner disk propagating toward the observer is partially absorbed and scattered by the dense, cool upper disk medium. In contrast, the thinner and hotter lower disk introduces negligible radiative obscuration, allowing unattenuated bright emission to reach the observer. This vertical flux discrepancy matches the features in the first row of Fig.~\ref{fig_hou}, including the downward shift of the photon ring and enhanced brightness in the lower half of the shadow image.

In summary, the MOG parameter \(\alpha\) dominantly regulates the amplitude and radial position of the intensity peak, while the viewing inclination angle \(\theta\) governs the spatial extent of accretion disk emission and induces asymmetric radiative modulation of the black hole shadow.

\section{Polarization imaging}\label{sec:5}
In this subsection, we investigate the polarimetric signatures of a Schwarzschild-MOG black hole under the Hou disk model framework, aiming to explain how the MOG parameter $\alpha$, intrinsic accretion dynamics, and viewing inclination angle $\theta$ jointly modulate the near-horizon polarization structure. For polarized imaging calculations, we exclusively adopt anisotropic radiation for the Hou disk model. Consistent with the setup in the preceding section, the accretion flow is prescribed to be infalling throughout all simulations.

\subsection{Covariant radiative transfer equation}
Under the WKB approximation, the propagation of light rays satisfies the radiative transfer equation:
\begin{equation}
k^\mu \nabla_\mu S^{\alpha\beta} = J^{\alpha\beta} + H^{\alpha\beta\mu\nu} S_{\mu\nu},\label{eq58}
\end{equation}
where $k^\mu$ is the wave vector of the light ray, $S^{\alpha\beta}$ is the polarization tensor used to characterize the  polarization state of the light, $J^{\alpha\beta}$ describes the emission properties of the radiation source, and $H^{\alpha\beta\mu\nu}$ characterizes the response of the propagation medium to light rays, which typically includes absorption effects and Faraday rotation effects.
For numerical solutions of the radiative transfer equation, one may refer to the open-source code Coport 1.0~\cite{huang2024coport}.

By using the gauge invariance of $S^{\alpha\beta}$, calculations can be carried out in a simple parallel transport frame. In this case, the covariant radiative transfer equation~\eqref{eq58} is decomposed into two parts. The first part describes gravitational effects,
\begin{equation}
k^\mu \nabla_\mu f^a = 0, \quad f^a k_a = 0,
\end{equation}
where $f^\mu$ is a normalized spacelike vector orthogonal to $k^\mu$. The second part reduces to the radiative transfer equation
\begin{equation}
\frac{d}{d\lambda} \mathbf{S} = \mathbf{R}(\chi) \mathbf{J} - \mathbf{R}(\chi) \mathbf{M} \mathbf{R}(-\chi) \mathbf{S},
\end{equation}
where
\begin{equation}
\mathbf{S} = 
\begin{pmatrix}
\mathcal{I} \\
\mathcal{Q} \\
\mathcal{U}  \\
\mathcal{V} 
\end{pmatrix}, \quad 
\mathbf{J} = \frac{1}{\nu^2}
\begin{pmatrix}
j_I \\
j_Q \\
j_U \\
j_V
\end{pmatrix}, \quad
\mathbf{M} = \nu
\begin{pmatrix}
a_I & a_Q & a_U & a_V \\
a_Q & a_I & r_V & -r_U \\
a_U & -r_V & a_I & r_Q \\
a_V & r_U & -r_Q & a_I
\end{pmatrix}.
\end{equation}
Note that $\mathbf{R}(\chi)$ is the rotation matrix accounting for the rotation between the synchrotron emission basis and the parallel-transported reference basis,
\begin{equation}
\mathbf{R}(\chi) = 
\begin{pmatrix}
1 & 0 & 0 & 0 \\
0 & \cos(2\chi) & -\sin(2\chi) & 0 \\
0 & \sin(2\chi) & \cos(2\chi) & 0 \\
0 & 0 & 0 & 1
\end{pmatrix}.\label{eq43}
\end{equation}
And $\chi$ is the rotation angle between the reference vector $f^\mu$ and the comoving magnetic field $b^\mu$ in the transverse subspace of the light ray,
\begin{equation}
\chi = \mathrm{sign}(\epsilon_{\mu\nu\alpha\beta} u^\mu f^\nu b^\rho k^\sigma) \arccos\left( \frac{P^{\mu\nu} f_\mu b_\nu}{\sqrt{(P^{\mu\nu} f_\mu f_\nu)(P^{\alpha\beta} b_\alpha b_\beta)}} \right),
\end{equation}
where $P^{\mu\nu}$ is the induced metric in the transverse subspace. In the observer's frame, the Stokes parameters need to be projected onto the observer's screen, which is accomplished via the rotation matrix~\eqref{eq43}, with the rotation angle given by 
\begin{equation}
\chi_o = \mathrm{sign}(\epsilon_{\mu\nu\rho\sigma} u^\mu f^\nu d^\rho k^\sigma) \arccos\Big( \frac{P^{\mu\nu} f_\mu d_\nu}{\sqrt{(P^{\mu\nu} f_\mu f_\nu) (P^{\alpha\beta} d_\alpha d_\beta)}} \Big).
\end{equation}
Here $d^\mu$ denotes the axis direction of the screen, and in this work we choose $d^\mu = -\partial_\theta^\mu$.

The projection results are then 
\begin{equation}
\mathcal{I_o} = \mathcal{I}, \quad 
\mathcal{Q_o} = \mathcal{Q} \cos \chi_o - \mathcal{U} \sin \chi_o\,, \quad
\mathcal{U_o} = \mathcal{Q} \sin \chi_o + \mathcal{U} \cos \chi_o\,, \quad \mathcal{V_o} = \mathcal{V},
\end{equation}
where the Stokes parameter $\mathcal{I_o}$ reflects the intensity of the light. Indeed, $\mathcal{V_o}$ also reflects the information about circular  polarization. If $\mathcal{V_o}$ is positive, it corresponds to left-handed circular polarization. Otherwise, it corresponds to right-handed circular polarization. $\mathcal{Q_o}$ and $\mathcal{U_o}$ reflect information about the electric field $\vec{E} = (E_x, E_y)$,
\begin{equation}
\mathcal{Q_o} = E_x^2 - E_y^2, \quad \mathcal{U_o} = 2E_x E_y.
\end{equation}
If $\mathcal{U_o}$ is positive, $E_x$ and $E_y$ have the same sign, and $\vec{E}$ lies in the first or third quadrant. If $\mathcal{U_o}$ is negative, $E_x$ and $E_y$ have opposite signs, and $\vec{E}$ lies in the second or fourth quadrant. The sign of $\mathcal{Q_o}$ reflects the relationship between $\vec{E}$ and $y = x$ (or $y = -x$). On the projection screen, the  polarization intensity corresponding to the polarization vector $\vec{f}$ and the EVPA become
\begin{equation}
P_o^2 = \mathcal{Q_o}^2 + \mathcal{U_o}^2, \quad 2\Phi_{\mathrm{EVPA}} =  \arctan\left( \mathcal{U_o}/\mathcal{Q_o} \right).\label{eq_f}
\end{equation}

\subsection{Numerical result}

\begin{figure}[!htb]
  \centering
  \begin{subfigure}{0.29\textwidth}
    \centering
    \includegraphics[width=\textwidth,height=0.3\textheight,keepaspectratio]{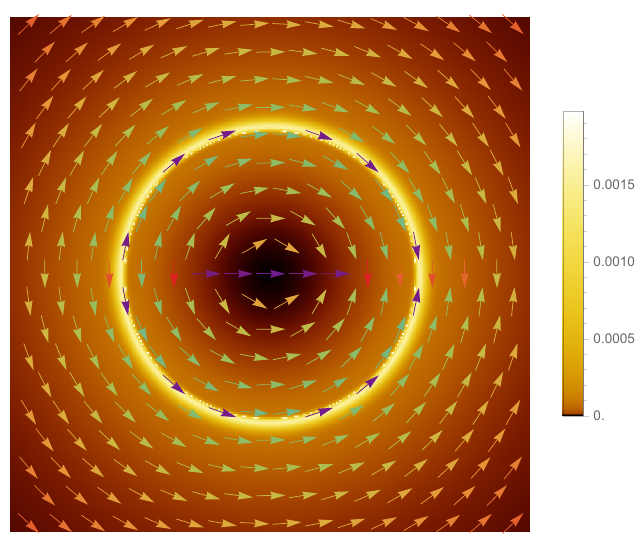}
    \caption{Stokes $\mathcal{I_o}$}
  \end{subfigure}
  \quad\quad
  \begin{subfigure}{0.29\textwidth}
    \centering
    \includegraphics[width=\textwidth,height=0.3\textheight,keepaspectratio]{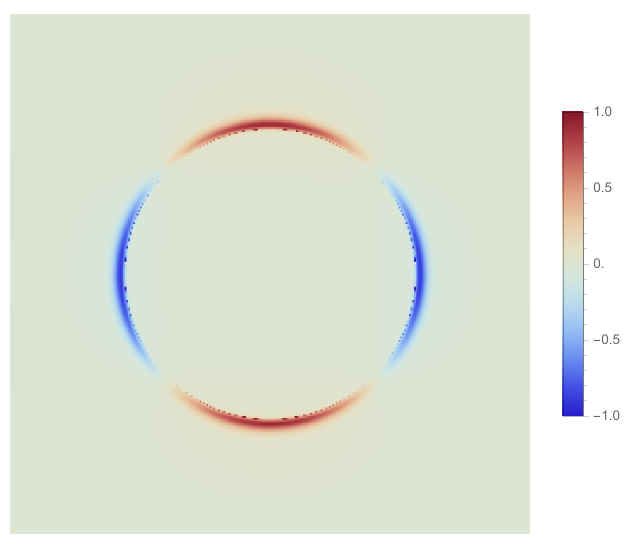}
   \caption{Stokes $\mathcal{Q_o}$}
  \end{subfigure}
  \quad\quad\quad
  \begin{subfigure}{0.29\textwidth}
    \centering
    \includegraphics[width=\textwidth,height=0.3\textheight,keepaspectratio]{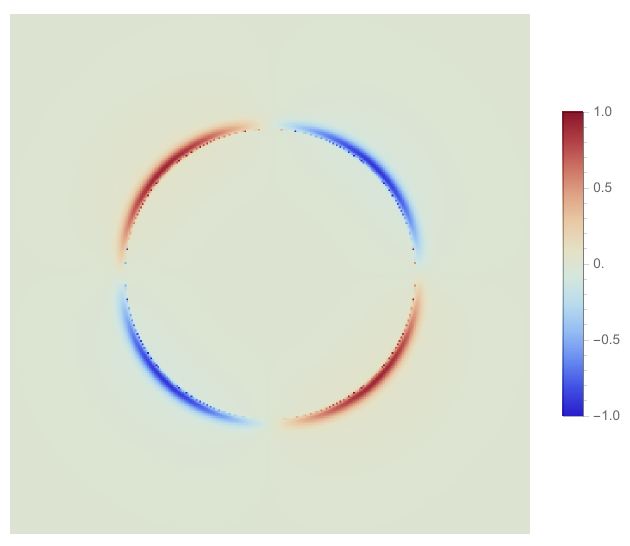}
    \caption{Stokes $\mathcal{U_o}$}
  \end{subfigure}
  \quad\quad
  \begin{subfigure}{0.29\textwidth}
    \centering
    \includegraphics[width=\textwidth,height=0.3\textheight,keepaspectratio]{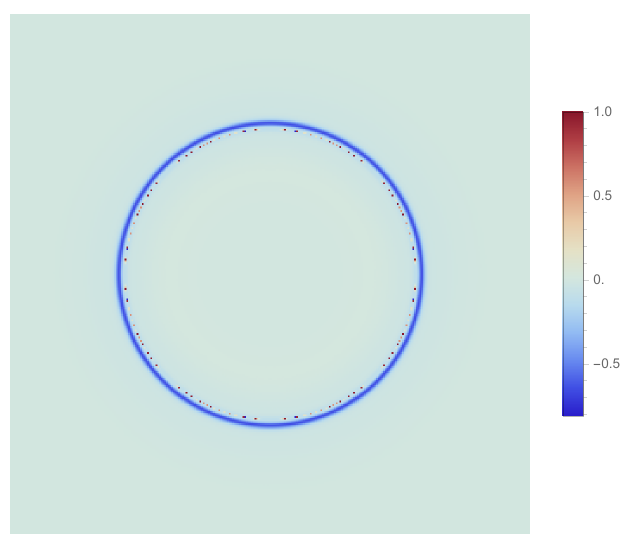}
    \caption{Stokes $\mathcal{V_o}$}
  \end{subfigure}
  \caption{
  Simulated distributions of the Stokes parameters $I_\ell$, $Q_\ell$, $U_\ell$, and $V_\ell$ for an infalling accretion flow within the Hou disk model. The parameters are fixed at $\alpha=0.99$, $\theta = 0.1^\circ$, and $\nu = 230\,\text{GHz}$.}
   \label{fig_IQUV}
\end{figure}
The polarization properties are shown in Fig.~\ref{fig_IQUV}. In the total intensity map $\mathcal{I_o}$, the overlaid arrows represent the polarization vector $\vec{f}$ evaluated via Eq.~\eqref{eq_f}, where the length and color depth denote the polarized intensity $P_o$, and the orientation traces $\Phi_{\mathrm{EVPA}}$. Since $\vec{f}$ is strictly orthogonal to the local magnetic field $\vec{b}$, the underlying magnetic field configuration can be inferred to be predominantly radial. The combination of $\mathcal{Q_o}$ and $\mathcal{U_o}$ qualitatively determines the orientation of the electric vector $\vec{E}$, while $\mathcal{V_o} < 0$ signifies circular polarization of the right handed. Furthermore, the linear and circular polarization signatures ($\mathcal{Q_o}$, $\mathcal{U_o}$, and $\mathcal{V_o}$) peak sharply within the higher-order lensing ring and decay rapidly toward zero away from this region. Given that the total intensity map $\mathcal{I_o}$ overlaid with polarization vectors encapsulates the essential polarimetric morphology, subsequent analysis will focus primarily on $\mathcal{I_o}$.

\begin{figure}[!htb]
  \centering
  \begin{subfigure}{0.29\textwidth}
    \centering
    \includegraphics[width=\textwidth,height=0.3\textheight,keepaspectratio]{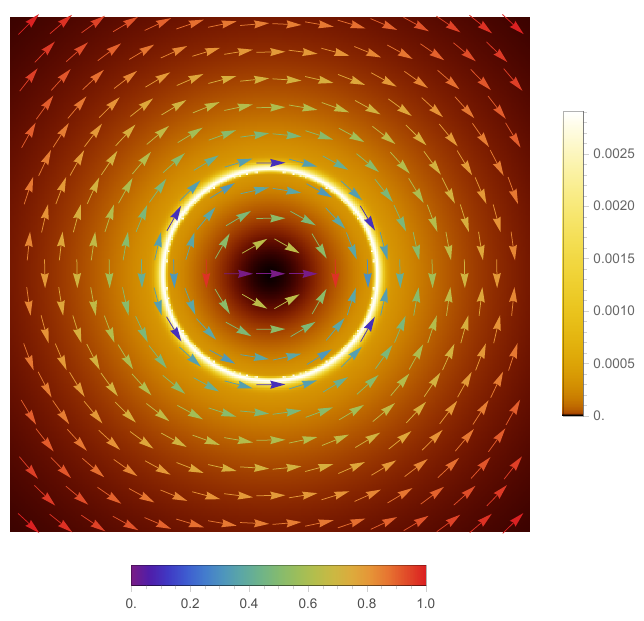}
    \caption{$\alpha=0.01$, $\theta=0.1^\circ$}
  \end{subfigure}
  \quad\quad
  \begin{subfigure}{0.29\textwidth}
    \centering
    \includegraphics[width=\textwidth,height=0.3\textheight,keepaspectratio]{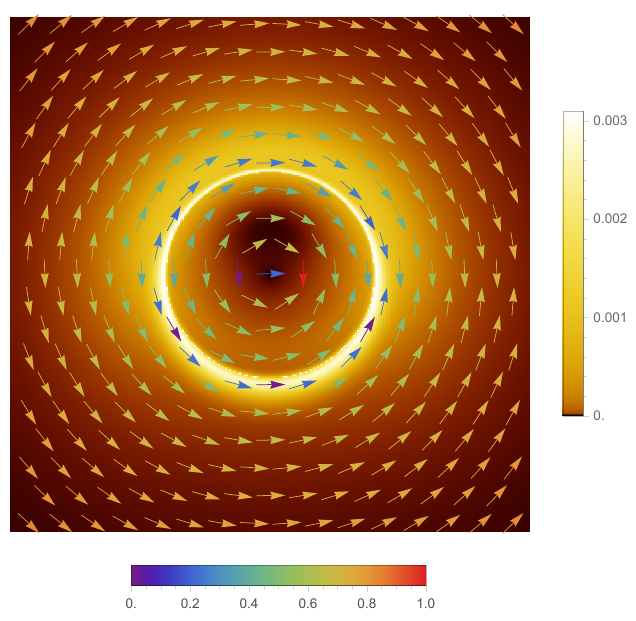}
   \caption{$\alpha=0.01$, $\theta=45^\circ$}
  \end{subfigure}
  \quad\quad
  \begin{subfigure}{0.29\textwidth}
    \centering
    \includegraphics[width=\textwidth,height=0.3\textheight,keepaspectratio]{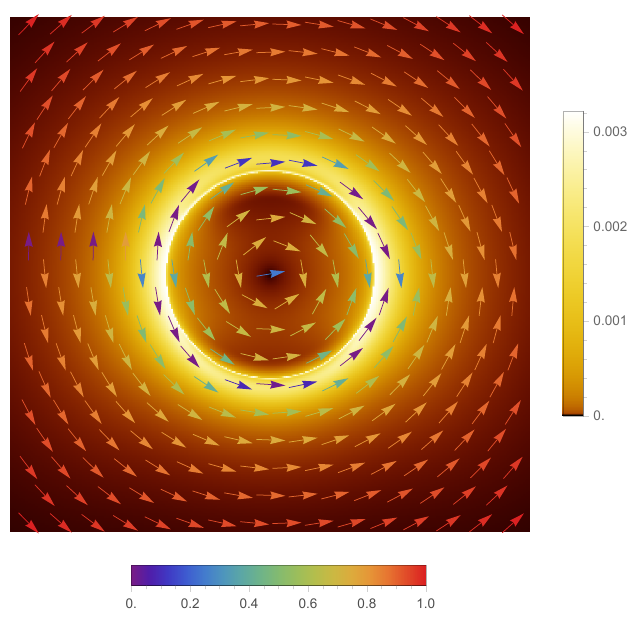}
  \caption{$\alpha=0.01$, $\theta=80^\circ$}
  \end{subfigure}
  \begin{subfigure}{0.29\textwidth}
    \centering
    \includegraphics[width=\textwidth,height=0.3\textheight,keepaspectratio]{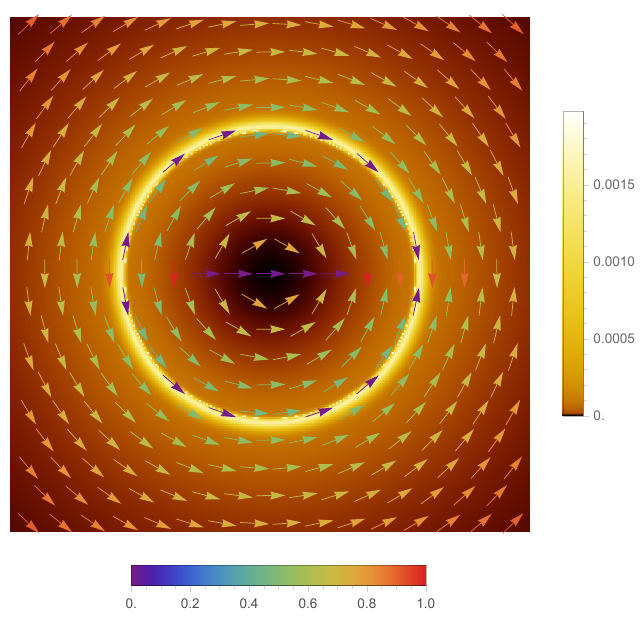}
    \caption{$\alpha=0.5$, $\theta=0.1^\circ$}
  \end{subfigure}
  \quad\quad
  \begin{subfigure}{0.29\textwidth}
    \centering
    \includegraphics[width=\textwidth,height=0.3\textheight,keepaspectratio]{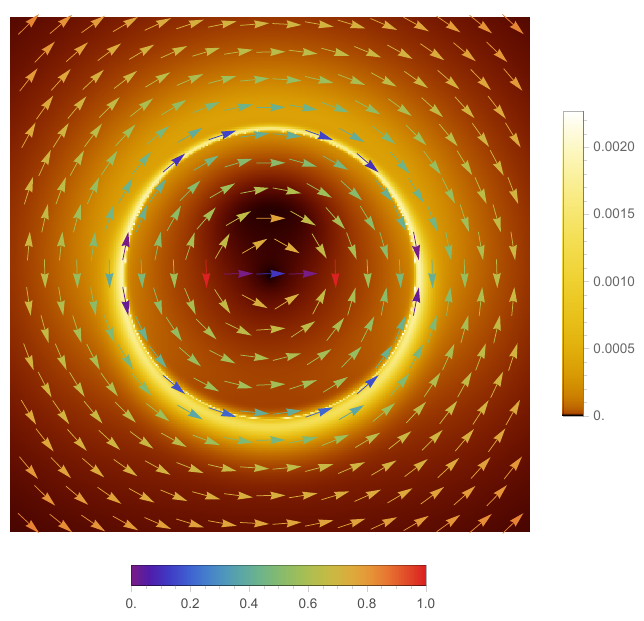}
    \caption{$\alpha=0.5$, $\theta=45^\circ$}
  \end{subfigure}
  \quad\quad
  \begin{subfigure}{0.29\textwidth}
    \centering
    \includegraphics[width=\textwidth,height=0.3\textheight,keepaspectratio]{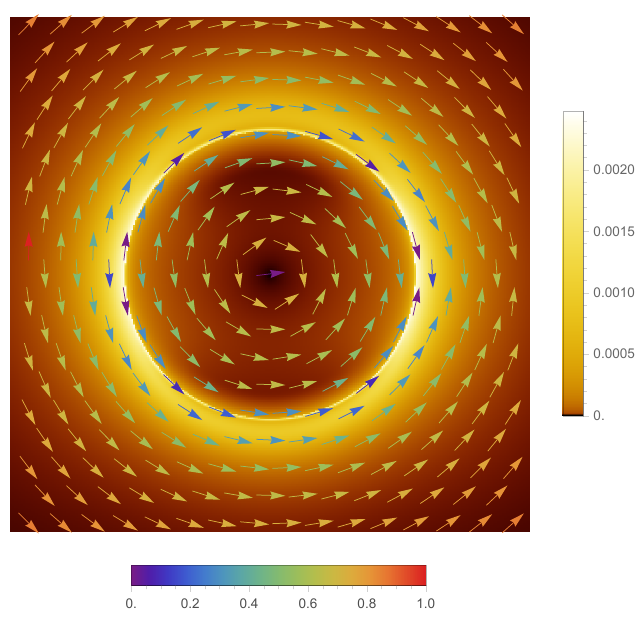}
    \caption{$\alpha=0.5$, $\theta=80^\circ$}
  \end{subfigure}
  \begin{subfigure}{0.29\textwidth}
    \centering
    \includegraphics[width=\textwidth,height=0.3\textheight,keepaspectratio]{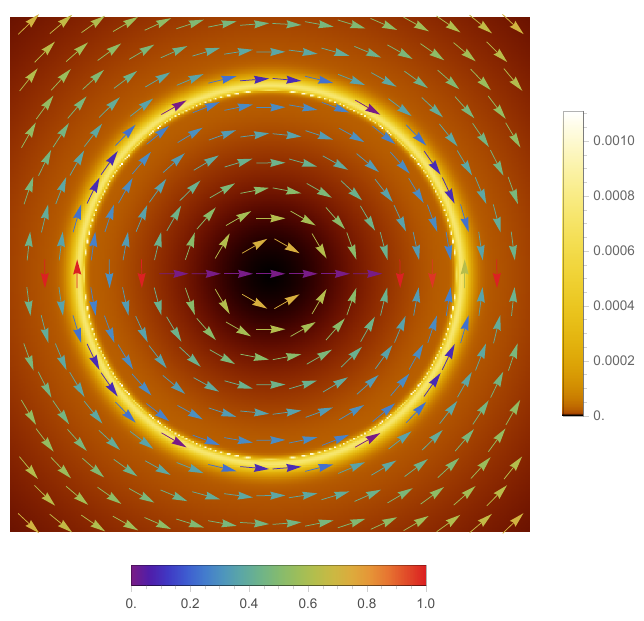}
    \caption{$\alpha=0.99$, $\theta=0.1^\circ$}
  \end{subfigure}
  \quad\quad
  \begin{subfigure}{0.29\textwidth}
    \centering
    \includegraphics[width=\textwidth,height=0.3\textheight,keepaspectratio]{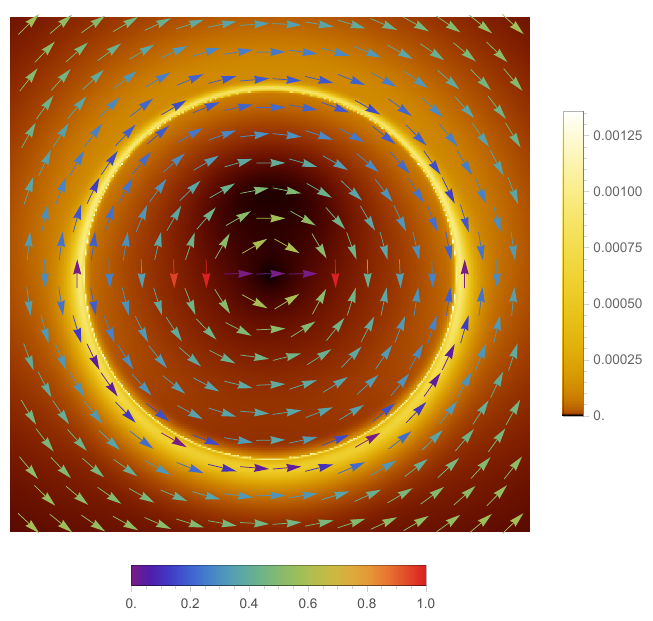}
    \caption{$\alpha=0.99$, $\theta=45^\circ$}
  \end{subfigure}
  \quad\quad
  \begin{subfigure}{0.29\textwidth}
    \centering
    \includegraphics[width=\textwidth,height=0.3\textheight,keepaspectratio]{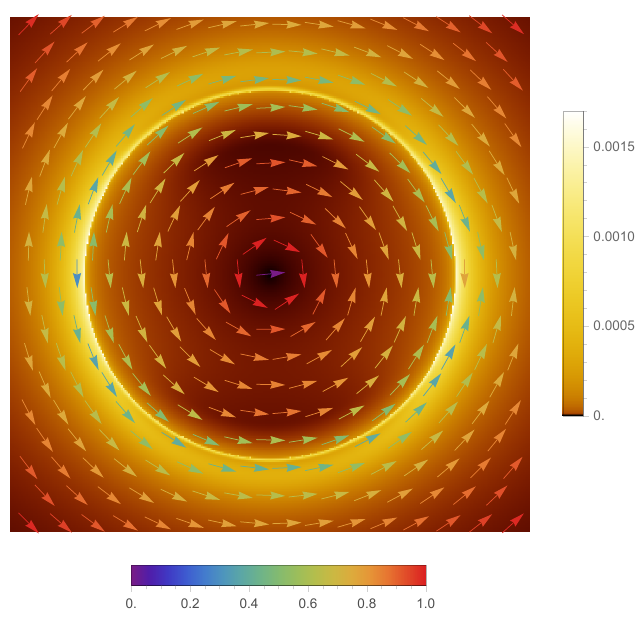}
    \caption{$\alpha=0.99$, $\theta=80^\circ$}
  \end{subfigure}
  \caption{Black hole shadow polarization images under the Hou disk model. The observer distance is fixed at $500M$, the field of view is $3^\circ$, and the observing frequency is $230\,\text{GHz}$.}
  \label{fig_polar}
\end{figure}
In Fig.~\ref{fig_polar}, the polarization signatures exhibit a pronounced dependence on the MOG parameter $\alpha$ and inclination angle $\theta$, with notable variations in both the morphology and the polarized flux density across different panels. As the parameter \(\alpha\) increases, the apparent size of the higher‑order images grows, accompanied by an increase in the number of polarization vectors within these images. It indicates that stronger gravitational coupling modifies the spatial distribution and dynamics of the accretion flow, enabling ordered magneti-field regions to form and persist over a more extended spatial scale. Consequently, regular linear-polarization signals can be detected on larger scales, which suggests that increasing \(\alpha\) facilitates the extension and stabilization of magnetic‑field configurations. Moreover, the observed mean polarized intensity rises with increasing viewing inclination. This is because of the enhanced contribution of transversely moving plasma, which produces synchrotron emission with a higher polarization degree. Our results demonstrate that a larger \(\alpha\) extends and stabilizes magnetic‑field structures, while a larger \(\theta\) boosts the mean polarized intensity. These two parameters  modulate the polarization properties of the system.

Overall, the EVPA pattern remains predominantly orthogonal to the radial direction (i.e., oriented azimuthally), while the polarization displays rapid spatial variations in the vicinity of the higher-order images. From top to bottom, as $\alpha$ increases, both the apparent size of the higher-order images and the central dark region (shadow) enlarge substantially. Concurrently, the overall orientation of the EVPA distribution undergoes noticeable shifts. Across each row, increasing the inclination angle $\theta$ causes the EVPA to deviate slightly from a purely azimuthal configuration, while the polarized intensity within the shadow region becomes increasingly prominent.

\begin{figure}[!htb]
  \centering
  \begin{subfigure}{0.35\textwidth}
    \centering
    \includegraphics[width=\textwidth,height=0.4\textheight,keepaspectratio]{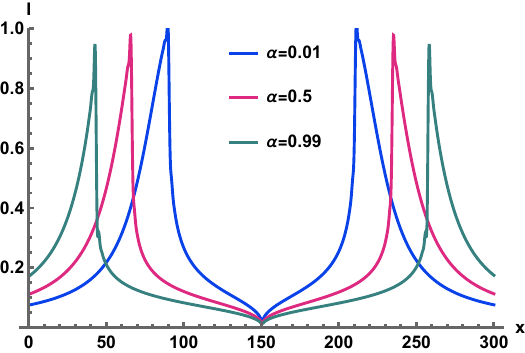}
    \caption{horizontal direction}
  \end{subfigure}
  \quad\quad
  \begin{subfigure}{0.35\textwidth}
    \centering
    \includegraphics[width=\textwidth,height=0.4\textheight,keepaspectratio]{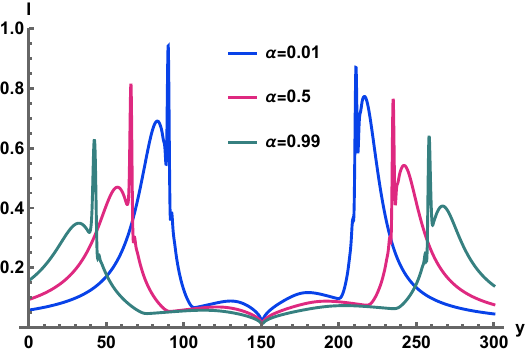}
   \caption{vertical direction}
  \end{subfigure}
   \caption{Intensity distribution profiles under different parameter values. The observer distance is fixed at $500M$, and the field-of-view angle is \(3^\circ\). Intensity distribution profiles for varying \(\alpha\): the left panel shows the horizontal intensity distribution profile, and the right panel shows the vertical intensity distribution profile. The viewing inclination is fixed at \(\theta=80^\circ\).}
  \label{fig_polarcurve}
\end{figure}
It is worth emphasizing that in geometrically thin disk models, the inner shadow coincides with the black hole event horizon, within which no polarized emission is present. In contrast, for the geometrically thick Hou disk considered here, gravitational lensing allows emission originating from regions above and below the equatorial plane to be projected onto the apparent horizon, thereby producing non-vanishing polarization vectors across the entire image plane. As shown in Fig.~\ref{fig_polarcurve}, the dominant intensity peaks correspond to the higher-order images generated in the Hou disk model. The peak amplitudes and positions vary significantly with $\alpha$, with the intensity decaying sharply to zero outside these peaks. These features demonstrate that the total flux distribution in the Hou disk model is predominantly governed by the higher-order lensed images. In contrast, phenomenological models typically maintain considerable primary image intensity well beyond the radii of the higher-order rings.

\section{Summary}\label{sec:6}
Black hole shadow imaging has become a frontier topic in both astrophysics and theoretical physics. The breakthrough observations of EHT have opened a new window for probing strong-field gravity and testing modified theories of gravity, significantly advancing numerical simulations of black hole imaging and its polarization signatures. In particular, effects from modified gravity  can modulate the luminosity and polarization features of black hole shadows, rendering them important observational diagnostics for distinguishing among competing gravitational models. Taking advantage of it, we adopted a geometrically thick and optically thin accretion flow as the radiation environment, and systematically investigated the shadow morphology and polarized imaging properties of the Schwarzschild-MOG black hole.

In this paper, we first reviewed the background of the Schwarzschild-MOG black hole and derived the geodesic equations for the photon‑sphere radius. Then we discussed two typical thick accretion‑flow models, namely the phenomenological model and the Hou disk model. By numerically solving the geodesic equations and radiative transfer equations, we obtained the corresponding optical images and polarization structures of the black hole. For unpolarized images, both the phenomenological model and the Hou disk model were considered. However, for polarimetric imaging, we only adopted the infalling-motion Hou disk model with the anisotropic emission.

For unpolarized shadow images, both isotropic and anisotropic emissions indicate that the MOG parameter primarily enhances the size and brightness of higher-order photon rings without altering their intrinsic morphology, while the inclination angle dominates the symmetry of the radiation field. Nearly face-on viewing geometries produce axisymmetric brightness distributions, whereas large inclination angles break the spatial symmetry of accretion emission, split the central dark shadow region, and broaden the transverse width of the photon ring. Compared with the phenomenological model, the Hou disk model exhibits notably suppressed off-equatorial radiative interference on the inner shadow, with higher-order images restricted to a narrower spatial regime, a behavior attributed to the reduced geometric thickness of the Hou disk under the conical approximation. Comparative analyses further confirmed that the anisotropic phenomenological model yields significantly brighter direct photon rings and more prominent higher-order imaging structures. At high inclinations, enhanced gravitational lensing and optical-depth projection effects in this model amplify the obscuration of the horizon contour by off-equatorial radiation, demonstrating the essential role of accretion geometry and emission anisotropy in shaping observational shadow signatures.

Polarization analyses performed for the Hou disk model reveal pronounced spatial localization of polarization signals. The EVPA patterns predominantly follow an azimuthal orientation with rapid spatial variations near the higher-order images. Notably, increasing the parameter \(\alpha\) substantially broadens both the shadow and the lensed photon rings alongside a global shift in EVPA orientation, while increasing the inclination \(\theta\) induces a slight deviation from pure azimuthal alignment and enhances polarized emission inside the shadow region.

\section*{Acknowledgments}
This work is supported by the Fundamental Research Funds for the Central Universities Project (Grant No.~2024IAIS-ZD009) and the National Natural Science Foundation of China (Grants No.~12505057 and No.~12547101).

\bibliography{ref} 

@article{oppenheimer1939continued,
  title={On Continued Gravitational Contraction},
  author={Oppenheimer, J. R. and Snyder, H.},
  journal={Physical Review},
  volume={56},
  number={5},
  pages={455--459},
  year={1939},
  doi={10.1103/PhysRev.56.455}
}

@article{penrose1965gravitational,
  title={Gravitational Collapse and Space-Time Singularities},
  author={Penrose, R.},
  journal={Physical Review Letters},
  volume={14},
  number={3},
  pages={57--59},
  year={1965},
  doi={10.1103/PhysRevLett.14.57}
}

@book{hawking1973large,
  title={The Large Scale Structure of Space-Time},
  author={Hawking, S. W. and Ellis, G. F. R.},
  publisher={Cambridge University Press},
  year={1973},
  doi={10.1017/CBO9780511524646}
}

@article{abbott2016observation,
  title={Observation of Gravitational Waves from a Binary Black Hole Merger},
  author={Abbott, B. P. and others},
  journal={Physical Review Letters},
  volume={116},
  number={6},
  pages={061102},
  year={2016},
  doi={10.1103/PhysRevLett.116.061102},
  eprint={1602.03837},
  eprinttype={arxiv},
  eprintclass={gr-qc}
}

@article{akiyama2019firstm87,
  title={First M87 Event Horizon Telescope Results. I. The Shadow of the Supermassive Black Hole},
  author={Akiyama, Kazunori and others},
  journal={The Astrophysical Journal Letters},
  volume={875},
  number={1},
  pages={L1},
  year={2019},
  doi={10.3847/2041-8213/ab0ec7},
  eprint={1906.11238},
  eprinttype={arxiv},
  eprintclass={astro-ph.GA}
}

@article{akiyama2022firstsgr,
  title={First Sagittarius A$^*$ Event Horizon Telescope Results. I. The Shadow of the Supermassive Black Hole in the Center of the Milky Way},
  author={Akiyama, Kazunori and others},
  journal={The Astrophysical Journal Letters},
  volume={930},
  number={2},
  pages={L12},
  year={2022},
  doi={10.3847/2041-8213/ac6674},
  eprint={2205.04245},
  eprinttype={arxiv},
  eprintclass={astro-ph.GA}
}

@article{akiyama2021firstm87vii,
  title={First M87 Event Horizon Telescope Results. VII. Polarization of the Ring},
  author={Akiyama, Kazunori and others},
  journal={The Astrophysical Journal Letters},
  volume={910},
  number={1},
  pages={L12},
  year={2021},
  doi={10.3847/2041-8213/abf2c4},
  eprint={2105.04383},
  eprinttype={arxiv},
  eprintclass={astro-ph.GA}
}

@article{akiyama2021firstm87viii,
  title={First M87 Event Horizon Telescope Results. VIII. Magnetic Field Structure near The Event Horizon},
  author={Akiyama, Kazunori and others},
  journal={The Astrophysical Journal Letters},
  volume={910},
  number={1},
  pages={L13},
  year={2021},
  doi={10.3847/2041-8213/abf3eb},
  eprint={2105.04386},
  eprinttype={arxiv},
  eprintclass={astro-ph.GA}
}

@article{akiyama2024firstsgrvii,
  title={First Sagittarius A$^*$ Event Horizon Telescope Results. VII. Polarization of the Ring},
  author={Akiyama, Kazunori and others},
  journal={The Astrophysical Journal Letters},
  volume={964},
  number={2},
  pages={L25},
  year={2024},
  doi={10.3847/2041-8213/ad2682},
  eprint={2405.05745},
  eprinttype={arxiv},
  eprintclass={astro-ph.GA}
}

@article{akiyama2024firstsgrviii,
  title={First Sagittarius A$^*$ Event Horizon Telescope results. VIII. Physical interpretation of the polarized ring},
  author={Akiyama, Kazunori and others},
  journal={The Astrophysical Journal Letters},
  volume={964},
  number={2},
  pages={L26},
  year={2024},
  publisher={The American Astronomical Society}
}

@article{psaltis2020gravitational,
  title={Gravitational test beyond the first post-Newtonian order with the shadow of the M87 black hole},
  author={Psaltis, Dimitrios and Medeiros, Lia and Christian, Pierre and {\"O}zel, Feryal and Akiyama, Kazunori and Alberdi, Antxon and Alef, Walter and Asada, Keiichi and Azulay, Rebecca and Ball, David and others},
  journal={Physical review letters},
  volume={125},
  number={14},
  pages={141104},
  year={2020},
  publisher={APS}
}

@article{kocherlakota2021constraints,
  title={Constraints on black-hole charges with the 2017 EHT observations of M87},
  author={Kocherlakota, Prashant and others},
  journal={Physical Review D},
  volume={103},
  number={10},
  pages={104047},
  year={2021},
  doi={10.1103/PhysRevD.103.104047},
  eprint={2105.09343},
  eprinttype={arxiv},
  eprintclass={gr-qc}
}

@article{Riess1998,
  author  = {Riess, A. G. and others},
  title   = {Observational Evidence from Supernovae for an Accelerating Universe and a Cosmological Constant},
  journal = {The Astronomical Journal},
  year    = {1998},
  volume  = {116},
  number  = {3},
  pages   = {1009--1038}
}

@article{Perlmutter1999,
  author  = {Perlmutter, S. and others},
  title   = {Measurements of $\Omega$ and $\Lambda$ from 42 High‑Redshift Supernovae},
  journal = {The Astrophysical Journal},
  year    = {1999},
  volume  = {517},
  number  = {2},
  pages   = {565--586}
}

@article{astier2012observational,
 title={Observational Evidence of the Accelerated Expansion of the Universe},
 author={Astier, Pierre and Pain, Reynald},
 journal={Comptes Rendus Physique},
 volume={13},
 pages={521--538},
 year={2012},
 publisher={Elsevier}
}

@article{rubin1970rotation,
  title={Rotation of the Andromeda nebula from a spectroscopic survey of emission regions},
  author={Rubin, Vera C and Ford Jr, W Kent},
  journal={Astrophysical Journal, vol. 159, p. 379},
  volume={159},
  pages={379},
  year={1970}
}

@article{rubin1980rotational,
  title={Rotational properties of 21 SC galaxies with a large range of luminosities and radii, from NGC 4605/R= 4kpc/to UGC 2885/R= 122 kpc},
  author={Rubin, Vera C and Ford Jr, W Kent and Thonnard, Norbert},
  journal={Astrophysical Journal, Part 1, vol. 238, June 1, 1980, p. 471-487.},
  volume={238},
  pages={471--487},
  year={1980}
}

@article{moffat2013mog,
 title={The MOG weak field approximation and observational test of galaxy rotation curves},
 author={Moffat, John W and Rahvar, Sohrab},
 journal={Monthly Notices of the Royal Astronomical Society},
 volume={436},
 number={2},
 pages={1439--1451},
 year={2013},
 publisher={Oxford University Press}
}

@article{ade2016planck,
 title={Planck 2015 results. XIII. Cosmological parameters},
 author={Ade, PAR and others},
 journal={Astronomy \& Astrophysics},
 volume={594},
 pages={A13},
 year={2016},
 publisher={EDP Sciences}
}

@article{Planck2020,
  author  = {Aghanim, N. and others (Planck Collaboration)},
  title   = {Planck 2018 results. VI. Cosmological parameters},
  journal = {Astronomy \& Astrophysics},
  year    = {2020},
  volume  = {641},
  pages   = {A6}
}

@article{DiValentino2021,
  author  = {Di Valentino, E. and Mena, O. and Pan, S. and Visinelli, L. and Yang, W. and Melchiorri, A. and Mota, D. F. and Riess, A. G. and Silk, J.},
  title   = {In the realm of the Hubble tension—a review of solutions},
  journal = {Classical and Quantum Gravity},
  year    = {2021},
  volume  = {38},
  number  = {15},
  pages   = {153001}
}

@article{zwicky1937masses,
  title={On the Masses of Nebulae and of Clusters of Nebulae},
  author={Zwicky, Fritz},
  journal={The Astrophysical Journal},
  volume={86},
  pages={217},
  year={1937}
}

@article{trimble1987existence,
 title={Existence and nature of dark matter in the universe},
 author={Trimble, Virginia},
 journal={Annual Review of Astronomy and Astrophysics},
 volume={25},
 number={1},
 pages={425--472},
 year={1987},
 publisher={Annual Reviews}
}

@article{riess2004type,
 title={Type Ia supernova discoveries at z> 1 from the Hubble Space Telescope: Evidence for past deceleration and constraints on dark energy evolution},
 author={Riess, Adam G and Strolger, Louis-Gregory and Tonry, John and Casertano, Stefano and Ferguson, Henry C and Mobasher, Bahram and Challis, Peter and Filippenko, Alexei V and Jha, Saurabh and Kirshner, Robert P and others},
 journal={The Astrophysical Journal},
 volume={607},
 number={2},
 pages={665--687},
 year={2004},
 publisher={American Astronomical Society}
}

@article{Copeland2006,
  author  = {Copeland, E. J. and Sami, M. and Tsujikawa, S.},
  title   = {Dynamics of dark energy},
  journal = {International Journal of Modern Physics D},
  year    = {2006},
  volume  = {15},
  number  = {11},
  pages   = {1753--1936}
}

@article{Bertone2005,
  title={Particle dark matter: Evidence, candidates and constraints},
  author={Bertone, Gianfranco and Hooper, Dan and Silk, Joseph},
  journal={Physics reports},
  volume={405},
  number={5-6},
  pages={279--390},
  year={2005},
  publisher={Elsevier}
}

@article{astier2006supernova,
 title={The Supernova Legacy Survey: measurement of $\Omega_M$, $\Omega_\Lambda$, and $w$ from the first year data set},
 author={Astier, Pierre and Guy, Julien and Regnault, Nicolas and Pain, Reynald and Aubourg, Eric and Balam, David and Basa, Stephane and Carlberg, Raymond G and Fabbro, Sylvain and Fouchez, Dominique and others},
 journal={Astronomy \& Astrophysics},
 volume={447},
 number={1},
 pages={31--48},
 year={2006},
 publisher={EDP Sciences}
}

@article{Arcadi2018,
  author  = {Arcadi, G. and Dutra, M. and Ghosh, P. and Lindner, M. and Mambrini, Y. and Pierre, M. and Profumo, S. and Queiroz, F. S.},
  title   = {The waning of the WIMP? A review of models, searches, and constraints},
  journal = {The European Physical Journal C},
  year    = {2018},
  volume  = {78},
  number  = {3},
  pages   = {203}
}

@article{Clifton2012,
  author  = {Clifton, Timothy and Ferreira, Pedro G. and Padilla, Antonio and Skordis, Constantinos},
  title   = {Modified gravity and cosmology},
  journal = {Phys. Rep.},
  volume  = {513},
  number  = {1-3},
  pages   = {1--189},
  year    = {2012}
}

@article{Nojiri2017,
  author  = {Nojiri, S. and Odintsov, S. D. and Oikonomou, V. K.},
  title   = {Modified gravity theories on a nutshell: Inflation, bounce and late-time evolution},
  journal = {Phys. Rep.},
  volume  = {692},
  pages   = {1--104},
  year    = {2017}
}

@article{Glavan2020,
  author  = {Glavan, Denis and Lin, Chunshan},
  title   = {Einstein-Gauss-Bonnet Gravity in Four-Dimensional Spacetime},
  journal = {Phys. Rev. Lett.},
  volume  = {124},
  number  = {8},
  pages   = {081301},
  year    = {2020}
}

@article{Sotiriou2010,
  author  = {Sotiriou, Thomas P. and Faraoni, Valerio},
  title   = {$f(R)$ theories of gravity},
  journal = {Rev. Mod. Phys.},
  volume  = {82},
  number  = {1},
  pages   = {451--497},
  year    = {2010}
}

@article{Harko2011,
  author  = {Harko, Tiberiu and Lobo, Francisco S. N. and Nojiri, Shin'ichi and Odintsov, Sergei D.},
  title   = {$f(R, T)$ gravity},
  journal = {Phys. Rev. D},
  volume  = {84},
  number  = {2},
  pages   = {024020},
  year    = {2011}
}

@book{Fujii2003,
  author    = {Fujii, Yasunori and Maeda, Kei-ichi},
  title     = {The Scalar-Tensor Theory of Gravitation},
  publisher = {Cambridge University Press},
  address   = {Cambridge, England},
  year      = {2003}
}

@article{Cai2016,
  author  = {Cai, Yi-Fu and Capozziello, Salvatore and De Laurentis, Mariafelicia and Saridakis, Emmanuel N.},
  title   = {$f(T)$ teleparallel gravity and cosmology},
  journal = {Rep. Prog. Phys.},
  volume  = {79},
  number  = {10},
  pages   = {106901},
  year    = {2016}
}

@article{Moffat2006,
  author  = {Moffat, J. W.},
  title   = {Scalar-tensor-vector gravity theory},
  journal = {J. Cosmol. Astropart. Phys.},
  volume  = {2006},
  number  = {03},
  pages   = {004},
  year    = {2006}
}

@article{Boboqambarova:2021cbf,
    author = "Boboqambarova, Madina and Turimov, Bobur and Abdujabbarov, Ahmadjon",
    title = "{Particle motion around Schwarzschild-MOG black hole}",
    journal = "Mod. Phys. Lett. A",
    volume = "38",
    number = "10n11",
    pages = "2350071",
    year = "2023",
    note = "[Erratum: Mod.Phys.Lett.A 38, 2350071 (2023)]"
}

@article{Moffat:2014aja,
    author = "Moffat, J. W.",
    title = "{Black Holes in Modified Gravity (MOG)}",
    journal = "Eur. Phys. J. C",
    volume = "75",
    number = "4",
    pages = "175",
    year = "2015"
}

@article{Brownstein2006,
  author  = {Brownstein, J. R. and Moffat, J. W.},
  title   = {Galaxy Rotation Curves Without Nonbaryonic Dark Matter},
  journal = {Astrophys. J.},
  volume  = {636},
  number  = {2},
  pages   = {721--741},
  year    = {2006}
}

@article{Moffat2014,
  author  = {Moffat, J. W. and Rahvar, S.},
  title   = {Moffat’s modified gravity (MOG) and the Bullet Cluster},
  journal = {Mon. Not. R. Astron. Soc.},
  volume  = {441},
  number  = {4},
  pages   = {3724--3732},
  year    = {2014}
}

@article{Moffat2013CMB,
  author  = {Moffat, J. W. and Toth, V. T.},
  title   = {Cosmic microwave background, the acoustic peaks and the modified gravity (MOG) theory},
  journal = {Mon. Not. R. Astron. Soc.},
  volume  = {432},
  number  = {2},
  pages   = {1003--1009},
  year    = {2013}
}

@article{Moffat2009,
  author  = {Moffat, J. W. and Toth, V. T.},
  title   = {The modified gravity theory tested with spherical and elliptical galaxies},
  journal = {Class. Quantum Grav.},
  volume  = {26},
  number  = {8},
  pages   = {085002},
  year    = {2009}
}

@article{Moffat2013Galaxies,
  author  = {Moffat, J. W. and Toth, V. T.},
  title   = {Modified Gravity: Cosmology without dark matter or Einstein's cosmological constant},
  journal = {Galaxies},
  volume  = {1},
  number  = {1},
  pages   = {65--82},
  year    = {2013}
}

@article{Lima2020KerrMOG,
  author  = {Lima Junior, Haroldo C. D. and Crispino, Lu{\'\i}s C. B. and Cunha, Pedro V. P. and Herdeiro, Carlos A. R.},
  title   = {Spinning black holes with a massive vector field: rotating MOG black holes},
  journal = {Eur. Phys. J. C},
  volume  = {80},
  number  = {11},
  pages   = {1036},
  year    = {2020}
}

@article{Sheykhi2018,
  author  = {Sheykhi, Ahmad},
  title   = {Thermodynamics and shadow of modified gravity (MOG) black holes},
  journal = {Phys. Rev. D},
  volume  = {98},
  number  = {2},
  pages   = {024040},
  year    = {2018}
}

@article{Sharif2017,
  author  = {Sharif, M. and Shahzadi, M.},
  title   = {Particle motion near charged black holes in modified gravity},
  journal = {Eur. Phys. J. C},
  volume  = {77},
  number  = {6},
  pages   = {363},
  year    = {2017}
}

@article{Manfredi2018,
  author  = {Manfredi, Luca and Mureika, Jonas and Moffat, John},
  title   = {Quasinormal modes of modified gravity (MOG) black holes},
  journal = {Phys. Rev. D},
  volume  = {98},
  number  = {2},
  pages   = {024014},
  year    = {2018}
}

@article{Liang2022,
  author  = {Liang, Dong and Shen, Yong and Zhang, Xiang},
  title   = {Quasinormal modes and ringdown of Schwarzschild-MOG black holes},
  journal = {Eur. Phys. J. C},
  volume  = {82},
  number  = {5},
  pages   = {431},
  year    = {2022}
}

@article{Mureika2016,
  author  = {Mureika, J. R. and Moffat, J. W. and Faizal, Mir},
  title   = {Black hole thermodynamics in modified gravity (MOG)},
  journal = {Phys. Rev. D},
  volume  = {94},
  number  = {4},
  pages   = {044060},
  year    = {2016}
}

@article{Pradhan2017,
  author  = {Pradhan, Parthapratim},
  title   = {Regular black holes in modified gravity (MOG) and particle acceleration},
  journal = {Eur. Phys. J. C},
  volume  = {77},
  number  = {3},
  pages   = {188},
  year    = {2017}
}

@article{Moffat2015Shadow,
  author  = {Moffat, J. W.},
  title   = {Modified gravity black holes and their shadow sizes in the Event Horizon Telescope},
  journal = {Eur. Phys. J. C},
  volume  = {75},
  number  = {3},
  pages   = {130},
  year    = {2015}
}

@article{Guo2021,
  author  = {Guo, Sen and Lu, Xiao and Li, Guan-Qiang},
  title   = {Shadow and photon sphere of regular black holes in MOG},
  journal = {Eur. Phys. J. C},
  volume  = {81},
  number  = {11},
  pages   = {102},
  year    = {2021}
}

@article{Abramowicz2013,
  author  = {Abramowicz, Marek A. and Fragile, P. Chris},
  title   = {Foundations of Black Hole Accretion Disk Theory},
  journal = {Living Rev. Relativ.},
  volume  = {16},
  pages   = {1},
  year    = {2013}
}

@article{Narayan1995,
  author  = {Narayan, Ramesh and Yi, Insu},
  title   = {Advection-dominated Accretion: Underfed Black Holes and Neutron Stars},
  journal = {Astrophys. J.},
  volume  = {452},
  pages   = {710--713},
  year    = {1995}
}

@article{EHT2022SgrATest,
  author  = {{Event Horizon Telescope Collaboration}},
  title   = {First {Sagittarius} {A}* {Event} {Horizon} {Telescope} Results. {VI}. {Testing} the Black Hole Metric},
  journal = {Astrophys. J. Lett.},
  volume  = {930},
  number  = {2},
  pages   = {L17},
  year    = {2022}
}

@article{Johnson2023ngEHT,
  author  = {Johnson, Michael D. and others},
  title   = {Key Science Goals for the Next-Generation {Event} {Horizon} {Telescope}},
  journal = {Galaxies},
  volume  = {11},
  number  = {3},
  pages   = {61},
  year    = {2023}
}

@article{broderick2009imaging,
  title={Imaging the black hole silhouette of M87: implications for jet formation and black hole spin},
  author={Broderick, Avery E and Loeb, Abraham},
  journal={The Astrophysical Journal},
  volume={697},
  number={2},
  pages={1164--1179},
  year={2009},
  publisher={The American Astronomical Society}
}

@article{connors1980polarization,
  title={Polarization features of X-ray radiation emitted near black holes.},
  author={Connors, Paul A and Piran, Tsvi and Stark, Richard F},
  journal={The Astrophysical Journal},
  volume={235},
  pages={224--244},
  year={1980}
}

@article{akiyama2021,
  author  = {Akiyama, Kazunori and others},
  collaboration = {Event Horizon Telescope},
  title   = {First {M87} {Event} {Horizon} {Telescope} Results. {VIII}. Magnetic Field Structure near The Event Horizon},
  journal = {The Astrophysical Journal Letters},
  volume  = {910},
  number  = {1},
  pages   = {L13},
  year    = {2021},
  doi     = {10.3847/2041-8213/abe461}
}

@article{Narayan2021,
  author  = {Narayan, Ramesh and Palumbo, Daniel C. M. and Johnson, Michael D. and Gelles, Zachary and Himwich, Elizabeth and Prather, Ben S. and Chael, Andrew and Yfantis, Angelos and Alexandrou, Michalakis},
  title   = {The Polarized Image of a Synchrotron-emitting Ring of Gas Orbiting a Black Hole},
  journal = {The Astrophysical Journal},
  volume  = {912},
  number  = {1},
  pages   = {35},
  year    = {2021},
  doi     = {10.3847/1538-4357/abf117}
}

@article{Beloborodov2002,
  author  = {Beloborodov, Andrei M.},
  title   = {Gravitational Bending of Light Near Compact Objects},
  journal = {The Astrophysical Journal},
  volume  = {566},
  number  = {2},
  pages   = {L85--L88},
  year    = {2002},
  doi     = {10.1086/339511}
}

@article{Gelles2021a,
  author  = {Gelles, Zachary and Johnson, Michael D. and Narayan, Ramesh},
  title   = {The Polarized Image of an Equatorial Accretion Disk around a Black Hole},
  journal = {The Astrophysical Journal},
  volume  = {914},
  number  = {2},
  pages   = {118},
  year    = {2021},
  doi     = {10.3847/1538-4357/abfc36}
}

@article{Gelles2021b,
  author  = {Gelles, Zachary and Palumbo, Daniel C. M. and Johnson, Michael D.},
  title   = {Polarized Images of Equatorial Rings around {Kerr} Black Holes},
  journal = {The Astrophysical Journal},
  volume  = {921},
  number  = {2},
  pages   = {165},
  year    = {2021},
  doi     = {10.3847/1538-4357/ac20cb}
}

@article{deliyski2023polarized,
  title={Polarized image of equatorial emission in horizonless spacetimes: Naked singularities},
  author={Deliyski, Valentin and Gyulchev, Galin and Nedkova, Petya and Yazadjiev, Stoytcho},
  journal={Physical Review D},
  volume={108},
  number={10},
  pages={104049},
  year={2023},
  publisher={APS}
}

@article{Qin2022,
  author  = {Qin, Xiao and Chen, Songbai and Jing, Jiliang},
  title   = {Polarized image of an equatorial emitting ring around a {4D} {Gauss-Bonnet} black hole},
  journal = {The European Physical Journal C},
  volume  = {82},
  number  = {9},
  pages   = {784},
  year    = {2022},
  doi     = {10.1140/epjc/s10052-022-10753-8}
}

@article{yang2026shadow,
  title={Shadow and polarization images of rotating black holes in Kalb-Ramond gravity illuminated by several thick accretion disks},
  author={Yang, Chen‑Yu and Ye, Huan and Zeng, Xiao‑Xiong},
  journal={The European Physical Journal C},
  volume={86},
  number={4},
  pages={342},
  year={2026},
  publisher={Springer},
  doi={10.1140/epjc/s10052-026-15584-5},
  eprint={2510.21229},
  archivePrefix={arXiv},
  primaryClass={gr-qc}
}

@article{zeng2026impact,
  title={The impact of cold dark matter halo on black hole shadow and polarization images under thick disk illumination},
  author={Zeng, Xiao‑Xiong and Yang, Chen‑Yu and He, Ke‑Jian and Li, Li‑Fang},
  journal={The European Physical Journal C},
  volume={86},
  number={6},
  pages={614},
  year={2026},
  publisher={Springer},
  doi={10.1140/epjc/s10052-026-15827-5},
  eprint={2603.14511},
  archivePrefix={arXiv},
  primaryClass={gr-qc}
}

@article{zeng2025observational,
  title={Observational signatures and polarized images of rotating charged black holes in Kalb‑Ramond gravity},
  author={He, Ke‑Jian and Yang, Chen‑Yu and Zeng, Xiao‑Xiong and Chang, Zheng‑Xue},
  journal={Chinese Physics C},
  volume={50},
  number={3},
  pages={035102},
  year={2026},
  doi={10.1088/1674‑1137/ad9837},
  eprint={2508.07393},
  archivePrefix={arXiv},
  primaryClass={gr-qc}
}

@article{chen2025polarized,
  title={Polarized image of a synchrotron-emitting ring in an Einstein-Maxwell-scalar theory},
  author={Chen, Yiqian and Cheng, Lang and Wang, Peng and Yang, Haitang},
  journal={Physical Review D},
  volume={112},
  number={2},
  pages={024044},
  year={2025},
  publisher={APS}
}

@article{shi2024polarized,
  title={Polarized image of a synchrotron emitting ring around a static hairy black hole in Horndeski theory},
  author={Shi, Han-Yu and Zhu, Tao},
  journal={The European Physical Journal C},
  volume={84},
  number={8},
  pages={814},
  year={2024},
  publisher={Springer}
}

@incollection{NovikovThorne1973,
  author    = {Novikov, Igor D. and Thorne, Kip S.},
  title     = {Astrophysics of black holes},
  booktitle = {Black Holes (Les Astres Occlus)},
  editor    = {Dewitt, C. and Dewitt, B. S.},
  pages     = {343--450},
  publisher = {Gordon and Breach},
  address   = {New York},
  year      = {1973}
}

@article{moscibrodzka2016general,
  title={General relativistic magnetohydrodynamical simulations of the jet in M 87},
  author={Mo{\'s}cibrodzka, Monika and Falcke, Heino and Shiokawa, Hotaka},
  journal={Astronomy \& Astrophysics},
  volume={586},
  pages={A38},
  year={2016},
  publisher={EDP Sciences}
}

@article{pu2016odyssey,
  title={Odyssey: a public GPU-based code for general relativistic radiative transfer in Kerr spacetime},
  author={Pu, Hung-Yi and Yun, Kiyun and Younsi, Ziri and Yoon, Suk-Jin},
  journal={The Astrophysical Journal},
  volume={820},
  number={2},
  pages={105},
  year={2016},
  publisher={The American Astronomical Society}
}

@article{yuan2003nonthermal,
  title={Nonthermal electrons in radiatively inefficient accretion flow models of Sagittarius A},
  author={Yuan, Feng and Quataert, Eliot and Narayan, Ramesh},
  journal={The Astrophysical Journal},
  volume={598},
  number={1},
  pages={301--312},
  year={2003}
}

@article{broderick2011evidence,
  title={Evidence for low black hole spin and physically motivated accretion models from millimeter-VLBI observations of Sagittarius A},
  author={Broderick, Avery E and Fish, Vincent L and Doeleman, Sheperd S and Loeb, Abraham},
  journal={The Astrophysical Journal},
  volume={735},
  number={2},
  pages={110},
  year={2011},
  publisher={The American Astronomical Society}
}

@article{Hou2024,
  author  = {Hou, Yiqian and Zhang, Zhenyu and Guo, Minyong and Chen, Bin},
  title   = {A new analytical model of magnetofluids surrounding rotating black holes},
  journal = {Journal of Cosmology and Astroparticle Physics},
  volume  = {2024},
  number  = {02},
  pages   = {030},
  year    = {2024},
  doi     = {10.1088/1475-7516/2024/02/030}
}

@article{Zhang2024,
  author  = {Zhang, Zhenyu and Hou, Yiqian and Guo, Minyong and Chen, Bin},
  title   = {Imaging thick accretion disks and jets surrounding black holes},
  journal = {Journal of Cosmology and Astroparticle Physics},
  volume  = {2024},
  number  = {05},
  pages   = {032},
  year    = {2024},
  doi     = {10.1088/1475-7516/2024/05/032}
}

@article{gold2020verification,
  title={Verification of radiative transfer schemes for the EHT},
  author={Gold, Roman and Broderick, Avery E and Younsi, Ziri and Fromm, Christian M and Gammie, Charles F and Mo{\'s}cibrodzka, Monika and Pu, Hung-Yi and Bronzwaer, Thomas and Davelaar, Jordy and Dexter, Jason and others},
  journal={The Astrophysical Journal},
  volume={897},
  number={2},
  pages={148},
  year={2020},
  publisher={The American Astronomical Society}
}

@article{leung2011numerical,
  title={Numerical calculation of magnetobremsstrahlung emission and absorption coefficients},
  author={Leung, Po Kin and Gammie, Charles F and Noble, Scott C},
  journal={The Astrophysical Journal},
  volume={737},
  number={1},
  pages={21},
  year={2011},
  publisher={The American Astronomical Society}
}

@article{mahadevan1996harmony,
  title={Harmony in electrons: Cyclotron and synchrotron emission by thermal electrons in a magnetic field},
  author={Mahadevan, Rohan and Narayan, Ramesh and Yi, Insu},
  journal={arXiv preprint astro-ph/9601073},
  year={1996}
}

@article{pu2016effects,
  title={The effects of accretion flow dynamics on the black hole shadow of Sagittarius A},
  author={Pu, Hung-Yi and Akiyama, Kazunori and Asada, Keiichi},
  journal={The Astrophysical Journal},
  volume={831},
  number={1},
  pages={4},
  year={2016},
  publisher={The American Astronomical Society}
}

@article{akiyama2019first5,
  title={First M87 Event Horizon Telescope Results. V. Physical Origin of the Asymmetric Ring},
  author={Akiyama, Kazunori and others},
  journal={The Astrophysical Journal Letters},
  volume={875},
  number={1},
  pages={L5},
  year={2019},
  doi={10.3847/2041-8213/ab0f43},
  eprint={1906.11242},
  eprinttype={arxiv},
  eprintclass={astro-ph.GA}
}

@article{akiyama2024event,
  title={Event Horizon Telescope. The polarised image of M\,87},
  author={Akiyama, Kazunori and others},
  journal={Astronomy \& Astrophysics},
  volume={681},
  pages={A79},
  year={2024},
  doi={10.1051/0004-6361/202347932}
}

@article{huang2024coport,
  title={Coport: a new public code for polarized radiative transfer in a covariant framework},
  author={Huang, Jiewei and Zheng, Liheng and Guo, Minyong and Chen, Bin},
  journal={Journal of Cosmology and Astroparticle Physics},
  volume={2024},
  number={11},
  pages={054},
  year={2024},
  publisher={IOP Publishing}
}


\end{document}